\documentclass[reprint, superscriptaddress, amsmath, amssymb, aps, prapplied]{revtex4-2}

\usepackage{graphicx}
\usepackage{siunitx}
\usepackage{booktabs}
\usepackage[siunitx, RPvoltages]{circuitikz}
\DeclareSIUnit\bar{bar}
\DeclareSIUnit\angstrom{\text {Å}}
\usepackage{hyperref}
\usepackage[caption = false]{subfig}

\begin{document}

\title{Parasitic Two-Level Systems in Merged-Element Transmon Qubits}

\author{Etienne Daum}
\email{contact author: etienne.daum@kit.edu}
\affiliation{Physikalisches Institut, Karlsruhe Institute of Technology, 76131 Karlsruhe, Germany}
\author{Benedikt Berlitz}
\affiliation{Physikalisches Institut, Karlsruhe Institute of Technology, 76131 Karlsruhe, Germany}
\author{Steffen Deck}
\affiliation{Physikalisches Institut, Karlsruhe Institute of Technology, 76131 Karlsruhe, Germany}
\author{Alexey V. Ustinov}
\affiliation{Physikalisches Institut, Karlsruhe Institute of Technology, 76131 Karlsruhe, Germany}
\author{Jürgen Lisenfeld}
\email{contact author: juergen.lisenfeld@kit.edu}
\affiliation{Physikalisches Institut, Karlsruhe Institute of Technology, 76131 Karlsruhe, Germany}

\date{\today}

\begin{abstract}
In conventional transmon qubits, decoherence is dominated by a large number of parasitic two-level systems (TLS) residing at the edges of the large area coplanar shunt capacitor and junction leads. Avoiding these defects by improvements in design, fabrication, and materials proved to be a significant challenge that so far led to limited progress. The merged-element transmon qubit (``mergemon''), a recently proposed paradigm shift in transmon design, attempts to address these issues by engineering the Josephson junction to act as its own shunt capacitor. With its energy mostly confined within the junctions, efforts required to improve qubit coherence can be concentrated on the junction barrier, a potentially easier to control interface compared with exposed circuit areas. Incorporating an additional aluminum deposition and oxidation into the \textit{in situ} bandaged Niemeyer-Dolan technique, we were able to fabricate flux-tunable mergemon qubits achieving mean $T_{1}$ relaxation times of up to $\SI{130}{\micro \second}$ ($Q \approx 3.3 \times 10^{6}$). TLS spectroscopy under applied strain and electric fields, together with systematic design variations, revealed that even for mergemon qubits — despite their significantly reduced footprint and increased junction barrier volume — careful design considerations are still essential to avoid $T_{1}$ limitations due to surface loss.
\end{abstract}

\maketitle

\section{Introduction}\label{sec:introduction}

\begin{figure*}
\centering
\subfloat{\includegraphics[width=0.22\textwidth]{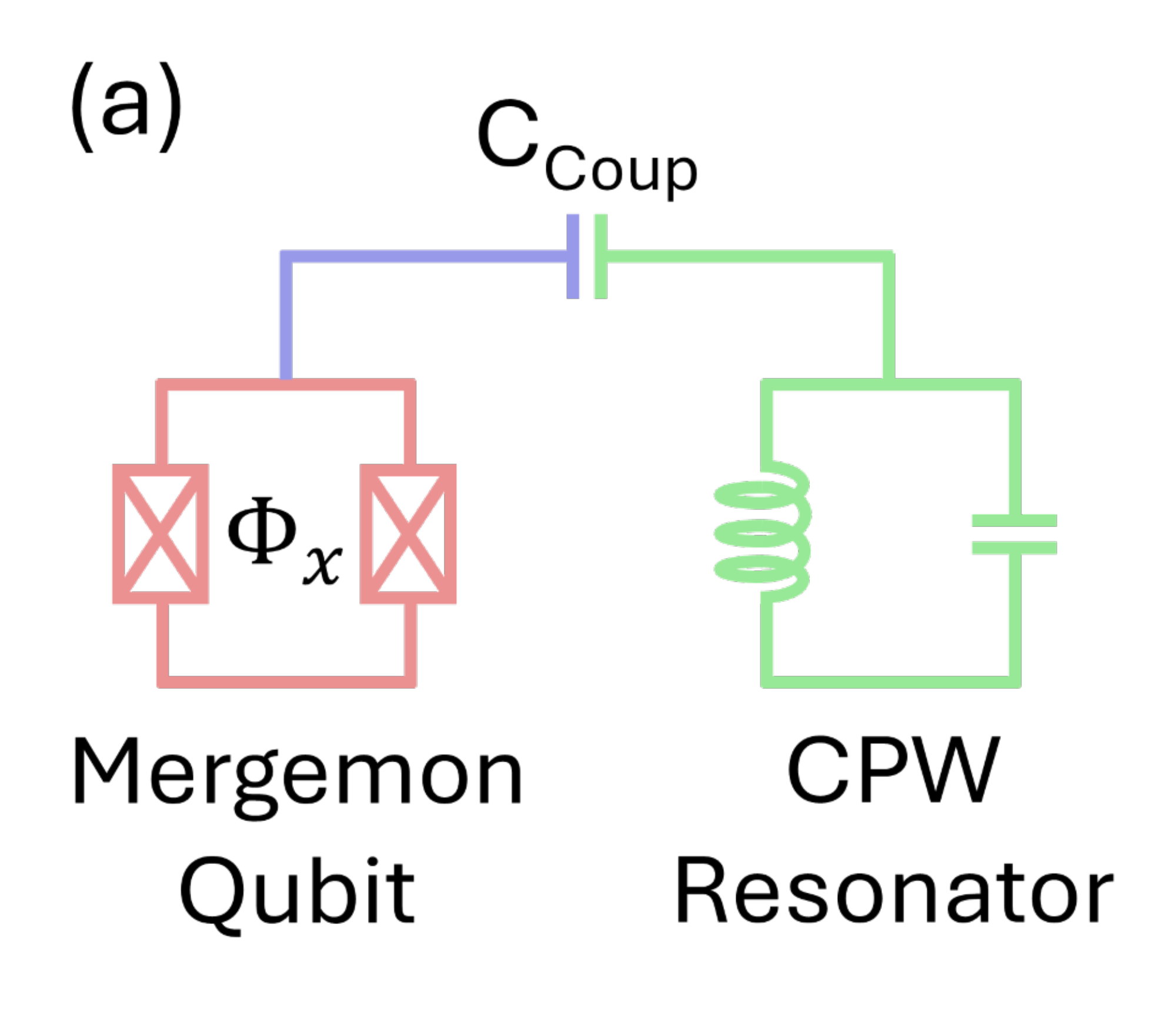}\label{fig:designa}}
\hfill
\subfloat{\includegraphics[width=0.325\textwidth]{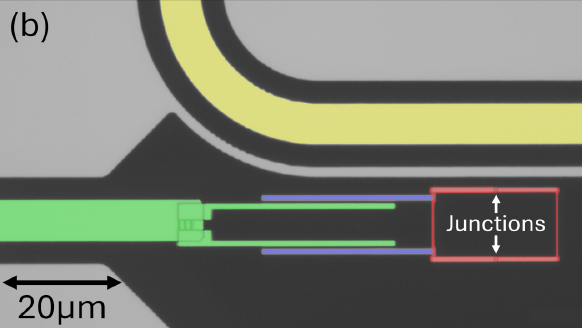}\label{fig:designb}}
\hfill
\subfloat{\includegraphics[width=0.437\textwidth]{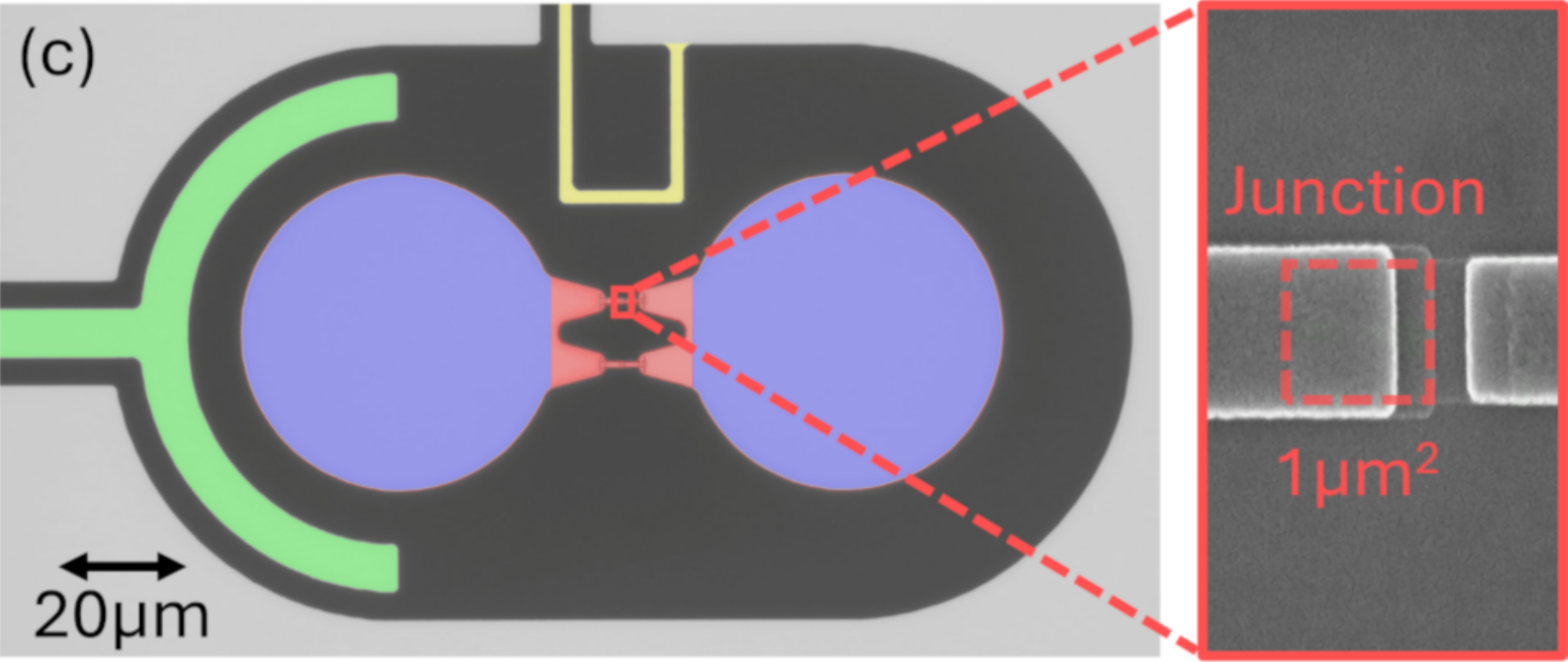}\label{fig:designc}}
\caption{(a) Mergemon-resonator circuit schematic. The mergemon consists of a loop of two identical, $\si{\micro \metre \squared}$-sized Josephson junctions that is capacitively coupled to the CPW readout resonator. The qubit frequency can be tuned by applying a flux $\Phi_{\mathrm{x}}$ through the loop. (b),(c) False-color optical micrographs of mergemon qubits AC (b) and BF1 (c), representing the two investigated mergemon design approaches A and B, respectively. While approach A aims to minimize the qubit footprint and maximize junction participation, approach B aims to minimize surface participation. Red: SQUID loop, blue: qubit islands and coupling to resonator, green: readout resonator, yellow: flux bias line, gray: ground plane, black: sapphire substrate. Inset: SEM image of the quadratic, $\SI{1}{\micro \metre \squared}$-sized Josephson junction employed in all designs of approach A and in design BF.}
\label{fig:design}
\end{figure*}

In recent years, the superconducting transmon qubit has become one of the most promising platforms for the realization of large scale quantum processors \cite{Google3,Google1,Google2,IBM}. It consists of a Josephson junction, providing the necessary nonlinearity, in parallel with a large-area, coplanar shunt capacitor, reducing the qubit's susceptibility to charge noise \cite{Koch}. While offering good coherence, easy coupling and readout, and a simple layout in comparison to other superconducting qubits, the gate fidelities of state-of-the-art transmon qubits are still too low to meet the requirements for practical quantum computation \cite{Kjaergaard,Google3}. Besides the employment of faster gates, more sophisticated quantum control techniques, and improved error correction codes, research primarily focuses on increasing transmon coherence. Substantial investigations identified parasitic two-level systems (TLS), strongly coupling to the transmon's electric fields via their dipole moments, as the dominating source of decoherence \cite{Lisenfeld_2019, Martinis2}. Significant effort is ongoing to mitigate these defects by improvements in design, fabrication, and materials \cite{Martinis, Bilmes, Tantalum1, Tantalum2, Tantalum3}.

With the main contribution of TLS originating from amorphous layers at the metal-air (MA), substrate-air (SA), and metal-substrate (MS) interfaces of the coplanar capacitor and junction leads \cite{Bilmes_2020,Wang_2015}, another strategy to address this issue is to simplify the transmon design by removing the coplanar shunt capacitor entirely. This recently proposed paradigm shift in transmon design, dubbed the merged-element transmon qubit (``mergemon''), engineers the Josephson junction to act as its own parallel shunt capacitor \cite{mergemonnotmamain, Mamin_2021}. By eliminating these large coplanar structures, the footprint of the mergemon qubit can be reduced by up to 2 orders of magnitude. The resulting localization of the electric field inside the Josephson junction inherently suppresses stray capacitive crosstalk to neighboring qubits as well as parasitic coupling to macroscopic box and antenna modes \cite{antennamode}. Consequently, mergemon-based quantum processors can be much more compact, given that the qubit readout and coupling circuitry is similarly scaled down \cite{kim2026tunablemodelesshybridizationfreecrosskerr}. Furthermore, the transition frequency of mergemon qubits could be less prone to fabrication-induced junction area fluctuations \cite{mergemonnotmamain}, leading to an improved qubit frequency targeting in large-scale quantum processors.

To meet the conventional requirements imposed on qubit frequency and Josephson energy to charging energy ratio ($E_{\mathrm{J}}/E_{\mathrm{C}}$), the area and thickness of the Josephson junction barrier need to be increased significantly. This leads to the introduction of a large number of junction TLS residing inside the amorphous barrier oxide \cite{PhysRevApplied.23.044054}. Because of the high electric fields inside the junctions, these TLS tend to be very strongly coupled, leading to a more pronounced impact on qubit performance than those residing at the MA, SA and MS interfaces. In the context of large-scale quantum processors, this currently limits the viability of the mergemon approach, as the presence of such strongly coupled TLS can effectively exclude entire qubits from operation. However, if future fabrication and material advances, like the employment of crystalline barriers \cite{Goswami_2022} or junction annealing \cite{Mamin_2021, Pappas_2024,wang2024precision}, succeed in sufficiently suppressing these defects, the mergemon approach has a large potential to outperform transmon-based processor architectures.

The high junction energy participation renders the mergemon qubit an ideal test bed for the study of junction TLS, as variations in junction fabrication, design, and post-processing techniques should show a pronounced impact on qubit coherence. This not only enables the development of targeted strategies for mitigating TLS-related loss, but may also offer deeper insight into their microscopic origins.

Here, we demonstrate mergemon qubits that achieve mean $T_{1}$ relaxation times of up to $\SI{130}{\micro \second}$ ($Q \approx 3.3 \times 10^{6}$), which is on par with conventional transmon qubits made from similar technology. We employ a fabrication technique, capable of realizing the needed, thicker-than-usual Josephson junction barriers without relying on hours long oxidations at extreme oxygen pressures (e.g., $\SI{4}{\hour}$ at $\SI{800}{\milli \bar}$ \cite{Mamin_2021}). Furthermore, via spectroscopy of individual TLS defects, we distinguish qubit energy relaxation due to surface- and junction TLS \cite{Lisenfeld_2019}. Our results indicate that the $T_{1}$ time of mergemon qubits, despite their significantly reduced footprint and increased junction barrier volume, can still be dominated by surface loss. However, successive optimization of the qubit geometry allowed us to minimize surface loss and present here mergemon qubits, which are no longer dominated by surface TLS, thereby demonstrating the potential of the mergemon design paradigm.

\section{Mergemon qubit design}\label{sec:des}

\begin{figure*}
\centering
\subfloat{\includegraphics[width=0.325\textwidth]{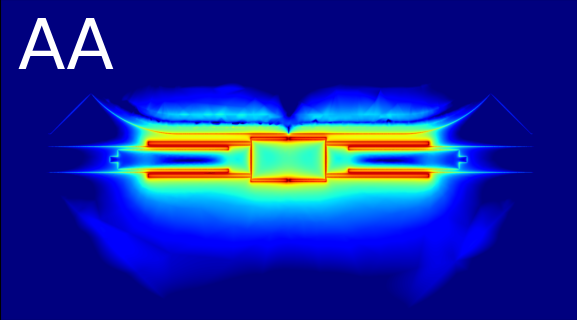}}
\hfill
\subfloat{\includegraphics[width=0.325\textwidth]{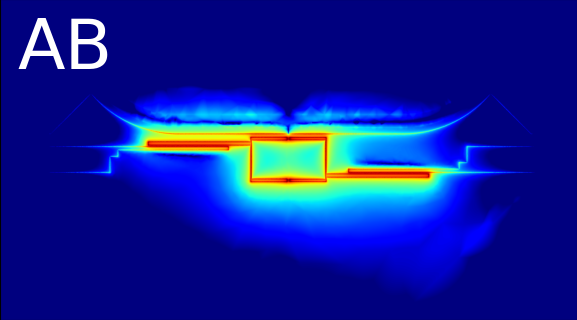}}
\hfill
\subfloat{\includegraphics[width=0.325\textwidth]{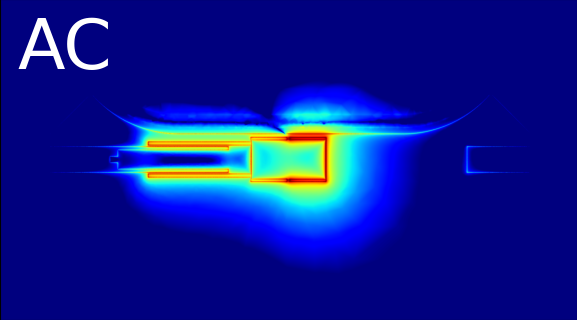}}
\vspace{-0.5em}
\subfloat{\includegraphics[width=0.325\textwidth]{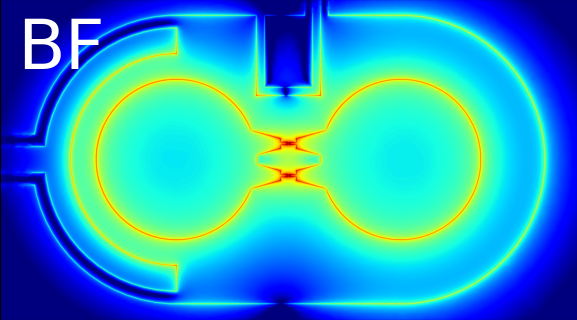}}
\hfill
\subfloat{\includegraphics[width=0.325\textwidth]{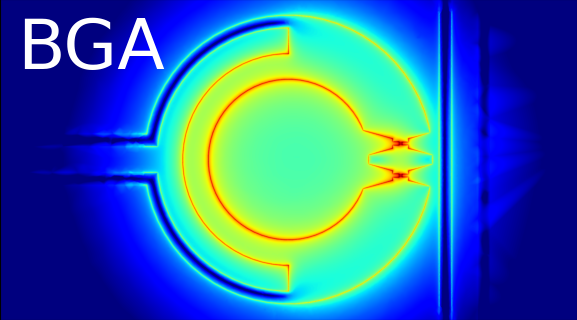}}
\hfill
\subfloat{\includegraphics[width=0.325\textwidth]{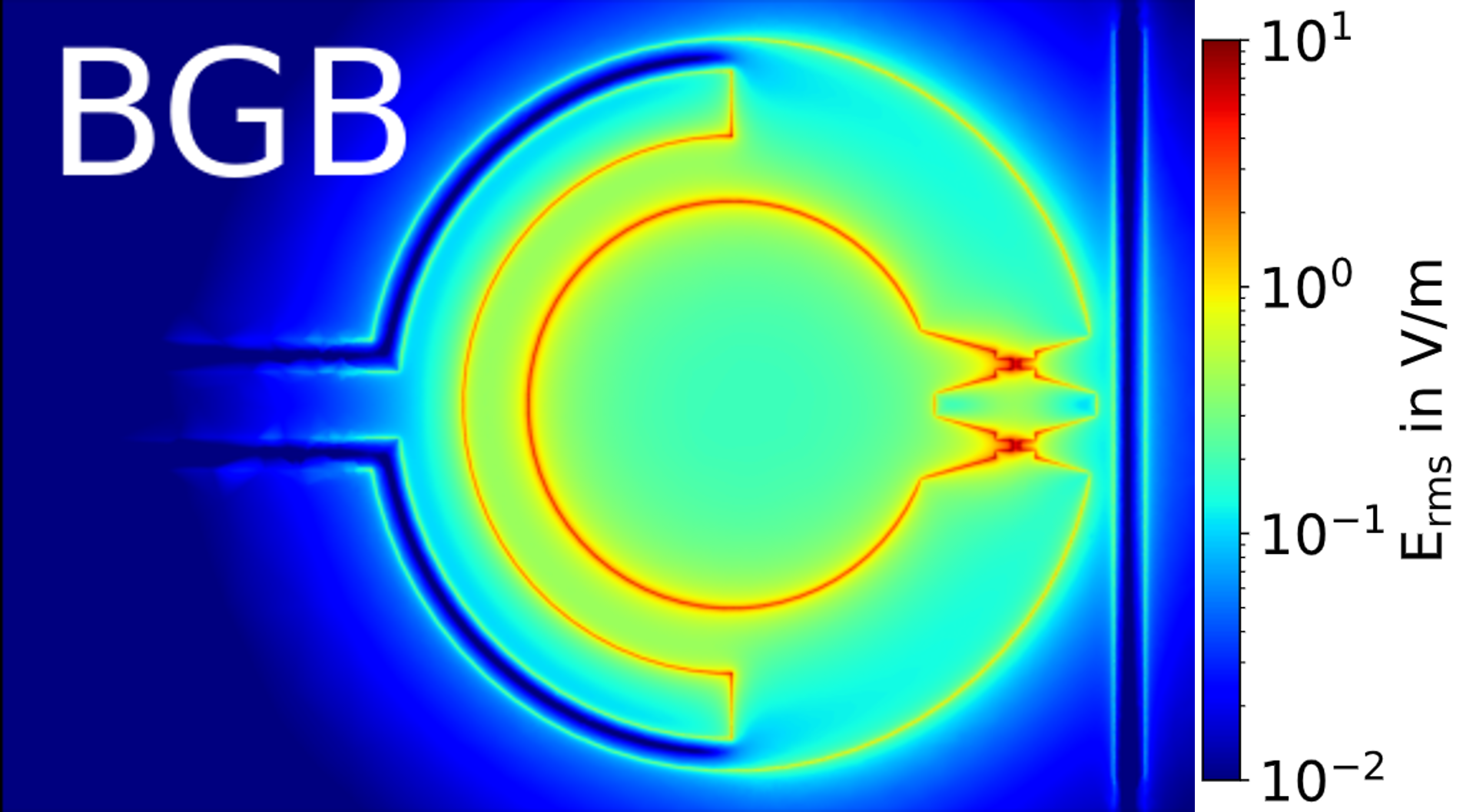}}
\caption{Simulated root-mean-square surface electric field strengths for all employed mergemon qubit designs at $\SI{5}{\giga \hertz}$ transition frequency. All plots utilize the same logarithmic color scale and window size of $\SI{180}{\micro \metre} \times \SI{100}{\micro \metre}$.}
\label{fig:fields}
\end{figure*}

The general idea of the mergemon approach is to engineer the Josephson junction to act as its own parallel shunt capacitor, thereby eliminating the need for an additional coplanar shunt capacitor as it is the case for the conventional transmon qubit. The circuitry of a flux-tunable mergemon qubit hence boils down to a loop of two identical, $\si{\micro \metre \squared}$-sized Josephson junctions (SQUID loop) with some capacitive structure coupling it to a readout resonator, as illustrated in Fig. \ref{fig:designa}.

Aiming for qubit frequencies between 4 and $\SI{8}{\giga \hertz}$, $E_{\mathrm{J}}/E_{\mathrm{C}}$ ratios exceeding 20 \cite{Koch}, and junction energy participation ratios $p_{\mathrm{JJ}} = C_{\mathrm{JJ,tot}}/C_{\mathrm{tot}} > 0.8$, the mergemon's Josephson junctions should provide a total capacitance between 50 and $\SI{100}{\femto \farad}$ and a total critical current between 20 to $\SI{80}{\nano \ampere}$. Utilizing $Al/AlO_{\mathrm{x}}/Al$ junctions, the barrier thickness $d_{\mathrm{JJ}}$ can be estimated to about 2 to $\SI{3}{\nano \metre}$ with a relative permittivity $\epsilon_{\mathrm{r}} \approx 10$. Assuming the junction to be a plate capacitor

\begin{equation}
C_{\mathrm{JJ}} = \frac{\epsilon_{0}\epsilon_{\mathrm{r}}A_{\mathrm{JJ}}}{d_{\mathrm{JJ}}},
\end{equation}

the combined junction area should be around $\SI{2}{\micro \metre \squared}$ to reach the targeted capacitance.

Starting from a conventional transmon junction ($A_{\mathrm{JJ}} \approx \SI{0.01}{\micro \metre \squared}$) and simply increasing its area by 2 orders of magnitude, the critical current, and thus the qubit frequency, would be shifted up into an unfavorable range. Therefore, the critical current density needs to be suppressed to about 1 to $\SI{4}{\ampere \per \centi \metre \squared}$, most readily achieved by increasing the junction barrier thickness. The required thickness is \textit{a priori} unknown and must be determined experimentally (see Sec. \ref{sec:fab}).

The mergemon design should further allow for dispersive qubit readout via capacitive coupling to a CPW resonator and for flux tuning of the qubit frequency by incorporating the junctions in a SQUID loop. Even though the mergemon's energy should be mostly confined within the junctions, the design of these additional structures is by no means trivial and can lead to surface participation ratios comparable to those of conventional transmon qubits for very small qubit footprints. To investigate the influence of different geometries on qubit performance, we employ six designs (see Fig. \ref{fig:fields}) with varying surface and junction participation ratios. These can be grouped into two approaches (A and B), exemplified in Fig. \ref{fig:designb} and \ref{fig:designc} by two representative designs.

Approach A aims to minimize the qubit footprint and maximize the junction participation $p_{\mathrm{JJ}}$. This is achieved through a floating design in which the qubit islands are effectively reduced to the about $\SI{1}{\micro \metre}$ wide leads connecting the junctions to form the SQUID loop. Capacitive coupling to the readout resonator is implemented by $\SI{1}{\micro \metre}$ wide antennas extending from the qubit as well as the resonator to form a coplanar capacitor with an effective coupling length of $\SI{25}{\micro \metre}$. All qubit structures are designed to be fabricated via additive lift-off. We employ three distinct qubit designs (AA, AB, AC) that follow this approach, differing in their coupling scheme and consequently their surface participation and footprints. Qubits AA and AB are coupled to both open ends of a $\lambda$/2-resonator, while qubit AC is coupled to the single open end of a $\lambda$/4-resonator. Qubits AB and AC each use two of the described coupling capacitors, while qubit AA employs four. All designs incorporate Josephson junctions with an area of about $\SI{1}{\micro \metre \squared}$. 

Approach B, on the other hand, aims to minimize the surface participation and defect formation at the cost of qubit footprint and $p_{\mathrm{JJ}}$. To dilute the surface fields and facilitate sample cleaning, the qubit islands are enlarged to $\SI{50}{\micro \metre}$ wide discs and connected by SQUID loops with tapered wires \cite{Martinis}. The extent of qubit structures fabricated via additive lift-off is reduced to $\SI{2}{\micro\metre}$-long leads adjacent to the junctions, potentially mitigating the formation of TLS in regions of strong qubit fields \cite{weeden2025statisticsstronglycoupleddefects}. Capacitive coupling to the single open end of a  $\lambda$/4-resonator is achieved by a clamp around one island. We employ three distinct qubit designs that follow this approach, with two of them being grounded (BGA, BGB) and one floating (BF). All three designs incorporate Josephson junctions of different sizes $\SI{1}{\micro \metre \squared}$ (BF), $\SI{1.25}{\micro \metre \squared}$ (BGA), and $\SI{1.5}{\micro \metre \squared}$ (BGB).

All designs utilize on-chip flux bias lines for fast qubit frequency tuning. The readout resonators are inductively coupled to a common transmission line and their resonance frequencies are centered around $\SI{7}{\giga \hertz}$ for approach A, and $\SI{9}{\giga \hertz}$ for approach B.

To support the proposed design rationale, we performed \textit{ANSYS HFSS} simulations (see App. \ref{app:simulation}) of all mergemon qubits, as well as one of our standard transmon designs \cite{lisenfeld2026mapping} to serve as a baseline for comparison. The resulting surface electric field distributions of the mergemon designs are plotted in Fig. \ref{fig:fields}. The extracted $p_{\mathrm{JJ}}$ and MS interface participation $p_{\mathrm{MS}}$, here serving as a proxy for the overall surface participation, as well as critical design parameters, are summarized in Tab. \ref{tab:designparameters} together with the results obtained by other groups on mergemon qubits (see App. \ref{app:simulation} for respective derivations). Notably, the $p_{\mathrm{JJ}}$ of our mergemon designs range from 0.83 to 0.96. In comparison to previously reported architectures with $p_{\mathrm{JJ}} > 0.93$, especially design approach B stretches the definition of a mergemon qubit a bit. However, with an at least 10 times higher junction participation than a conventional transmon qubit ($p_{\mathrm{JJ}} = 0.07$ in our case), we argue that these designs still follow the mergemon paradigm. The $p_{\mathrm{MS}}$ cover a broad range from $6.5 \times 10^{-5}$ to $3.6 \times 10^{-4}$, allowing for the systematic study of surface loss in mergemon qubits. As expected, the $p_{\mathrm{MS}}$ of approach B is about 2 to 6 times lower compared with approach A at the cost of up to 13\% less junction participation and a sevenfold increase in qubit footprint. Furthermore, despite a roughly 17-fold reduction in footprint, design AA exhibits a higher $p_{\mathrm{MS}}$ than our standard transmon. It shall be noted that the remaining energy, not captured by $p_{\mathrm{MS}}$ and $p_{\mathrm{JJ}}$, is predominantly stored in the vacuum and substrate, both of which have considerably lower dielectric loss tangents and therefore contribute minimally to the overall dissipation \cite{Wang_2015}.

\begin{table*}
\centering
\caption{Mergemon qubit design parameters and simulation results (see App. \ref{app:simulation} for exact derivations). Here $p_{\mathrm{JJ}}$ is the junction energy participation ratio, $p_{\mathrm{MS}}$ is the MS interface energy participation ratio, $p_{\mathrm{S}}$ is the substrate energy participation ratio, $C_{\mathrm{JJ}}$ is the calculated capacitance of a single Josephson junction, $C_{\mathrm{geom}}$ is the geometric qubit capacitance originating from the islands and junction leads, and $A_{\mathrm{JJ}}$ is the area of a single Josephson junction. For comparison, we further provide the results obtained for one of our standard transmon qubits (Transmon), and the results obtained by other groups on mergemon qubits (Zhao, Mamin).}
\label{tab:designparameters}
    \begin{tabular}{c c c c c c c c}
        \toprule
        Qubit & $p_{\mathrm{JJ}}$ & $p_{\mathrm{MS}}$ & $p_{\mathrm{S}}$ & $C_{\mathrm{JJ}}$ in $\si{\femto \farad}$ & $C_{\mathrm{geom}}$ in $\si{\femto \farad}$ & $A_{\mathrm{JJ}}$ in $\si{\micro \metre \squared}$ & Footprint in $\si{\micro \metre \squared}$\\
        \midrule
        AA & 0.92 & 3.63 $\times 10^{-4}$ & 0.070 & 28.5 & 4.76 & 1 & 1200 \\
        AB & 0.95 & 2.01 $\times 10^{-4}$ & 0.047 & 28.5 & 3.13 & 1 & 1200 \\
        AC & 0.96 & 1.54 $\times 10^{-4}$ & 0.032 & 28.5 & 2.07 & 1 & 800 \\
        \midrule
        BF & 0.89 & 6.53 $\times 10^{-5}$ & 0.098 & 34 & 8.20 & 1 & 6000 \\
        BGA & 0.83 & 9.80 $\times 10^{-5}$ & 0.157 & 36.5 & 15.22 & 1.25 & 3500 \\
        BGB & 0.86 & 7.85 $\times 10^{-5}$ & 0.125 & 48 & 15.24 & 1.5 & 3500 \\
        \midrule
        Transmon \cite{lisenfeld2026mapping} & 0.07 & 2.67 $\times 10^{-4}$ & 0.842 & 3.57 & 89.21 & 0.08 & 20000 \\
        \midrule
        Zhao et al. \cite{mergemonnotmamain} & 0.995 & - & - & 57 & - & 8 & - \\
        Mamin et al. \cite{Mamin_2021} & 0.93 & - & - & 62 & 5 & 1.4 & - \\
        \bottomrule
    \end{tabular}
\end{table*}

\section{Fabrication}\label{sec:fab}

In the first fabrication step, all structures with a minimum feature size of $\SI{2}{\micro \metre}$ were etched into an aluminum ground plane on a sapphire substrate utilizing optical lithography. In design approach A, this etching step excluded the entire qubit structures and was performed using an inductively coupled plasma etcher. In contrast, for approach B we used a wet etch and excluded only the Josephson junctions and their adjacent $\SI{2}{\micro\metre}$ long leads.

Subsequently, the remaining Josephson junctions and leads were fabricated using a modified version of the \textit{in-situ} bandaged Niemeyer-Dolan technique that avoids the formation of stray junctions \cite{Bilmes}. To realize the thicker-than-usual junction barriers, the junction bottom electrodes were first oxidized at a static pressure of $\SI{130}{\milli \bar}$ for $\SI{20}{\minute}$. To further thicken the barrier, an additional aluminum layer of less than $\SI{1}{\nano \metre}$ in thickness was deposited and equally oxidized.

\begin{figure}
\centering
\includegraphics[width = \columnwidth]{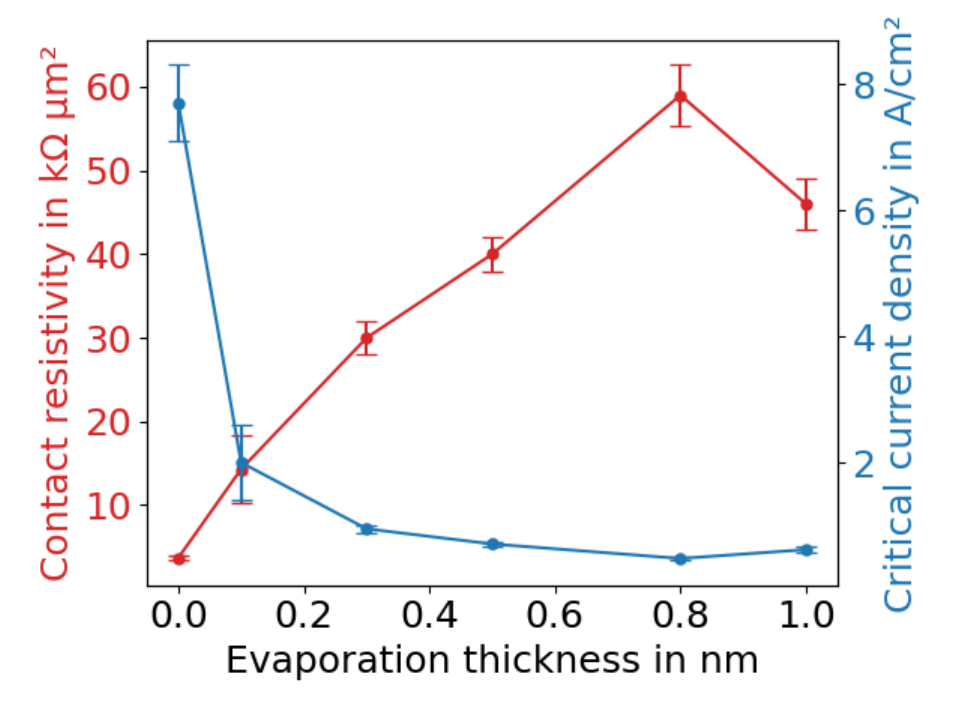}
\caption{Mean contact resistivity and derived critical current density of a total of 1400 measured SQUID loops with varying evaporation thicknesses of the second aluminum deposition. Transport measurements were performed directly after fabrication at room temperature. Critical current densities are derived from contact resistivities utilizing the Ambegaokar-Baratoff relation \cite{ambegaokar} and assume a superconducting energy gap of $\SI{182}{\micro \electronvolt}$ at $\SI{30}{\milli \kelvin}$. Junction areas were determined using SEM imaging.}
\label{fig:criticalcurrent}
\end{figure}

Figure \ref{fig:criticalcurrent} shows measurements of the room-temperature contact resistivities and calculated critical current densities of SQUID loops fabricated with different thicknesses of the additional aluminum layer. In contrast to the expected exponential dependency, we observe a linear increase in contact resistivity from 3.7 to $\SI{59}{\kilo \ohm \micro \metre \squared}$ below $\SI{0.8}{\nano \metre}$ layer thickness followed by a drop-off to $\SI{46}{\kilo \ohm \micro \metre \squared}$ at $\SI{1}{\nano \metre}$. While part of this behavior could be attributed to thickness fluctuations due to the evaporation system working at its limit of precision, it could also indicate an increasingly insufficient oxidation of the additional aluminum layer at higher thicknesses.

The corresponding critical current densities range from 0.48 to $\SI{7.7}{\ampere \per \centi \metre \squared}$. Accounting for a reduction in critical current density over time due to junction aging, an optimal working point, in accordance with the values derived in Sec. \ref{sec:des}, is reached for $\SI{0.1}{\nano \metre}$ evaporation thickness with a critical current density of $\SI{2}{\ampere \per \centi \metre \squared}$. While all mergemons characterized in Sec. \ref{sec:qubit} were fabricated with this layer thickness, for qubits following design approach B the oxidation pressure was lowered to $\SI{25}{\milli \bar}$ to reach higher $E_{\mathrm{J}}/E_{\mathrm{C}}$ ratios.

\begin{table*}[!htb]
\centering
\caption{Mergemon parameters obtained from qubit spectroscopy. Here, $f_{\mathrm{q}}$ is the qubit frequency at zero flux, $E_{\mathrm{C}}$ is the charging energy, $E_{\mathrm{J}}$ is the Josephson energy, $J_{\mathrm{C}}$ is the critical current density, $C_{\mathrm{JJ}}$ is the capacitance of a single Josephson junction, $T_{1}$ is the time-averaged relaxation time measured at the qubit frequency $f_{\mathrm{q,T_{1}}}$ (see App. Tab. \ref{tab:purcell}), $Q_{\mathrm{q}}$ is the corresponding quality factor, $g_{0}/2\pi$ is the qubit-resonator coupling strength at zero flux, and $T_{1}^{\mathrm{Purcell}}$ is the limitation due to Purcell loss at the qubit frequency $f_{\mathrm{q,T_{1}}}$. The uncertainties of the mean $T_{1}$ times and corresponding quality factors are given by the respective standard deviations.}
\label{tab:qubitparameters}
    \begin{tabular}{c c c c c c c c c c}
        \toprule
        Qubit & $f_{\mathrm{q}}$ in $\si{\giga \hertz}$ & $E_{\mathrm{C}}$ in $\si{\mega \hertz}$ & $E_{\mathrm{J}}/E_{\mathrm{C}}$ & $J_{\mathrm{C}}$ in $\si{\ampere \per \centi \metre \squared}$ & $C_{\mathrm{JJ}}$ in $\si{\femto \farad}$ & $T_{1}$ in $\si{\micro \second}$ & $Q_{\mathrm{q}} \times 10^{6}$ & $g_{0}/2\pi$ in $\si{\mega \hertz}$ & $T_{1}^{\mathrm{Purcell}}$ in $\si{\micro \second}$\\
        \midrule
        AA & 4.785 & 318 & 32.2 & 1.05 & 28.1 & 13 $\pm$ 5 & 0.36 $\pm$ 0.14 & 63 & 137 \\
        AB1 & 5.004 & 342 & 30.6 & 1.07 & 26.7 & 19 $\pm$ 4 & 0.54 $\pm$ 0.11 & 35 & 575 \\
        AB2 & 4.926 & 302 & 37.4 & 1.15 & 30.5 & 22 $\pm$ 3 & 0.53 $\pm$ 0.09 & 35 & 575 \\
        AC & 5.033 & 321 & 34.4 & 1.13 & 29.1 & 25 $\pm$ 8 & 0.73 $\pm$ 0.22 & 21 & 4000 \\
        \midrule
        BF1 & 7.567 & 258 & 115.1 & 3 & 33.5 & 39 $\pm$ 6 & 1.27 $\pm$ 0.19 & 62 & 447 \\
        BF2 & 7.502 & 246 & 124 & 3.1 & 35.3 & 49 $\pm$ 8 & 1.7 $\pm$ 0.3 & 64 & 555 \\
        BF3 & 7.688 & 251 & 125.1 & 3.2 & 34.5 & 52 $\pm$ 8 & 2 $\pm$ 0.3 & 66 & 375 \\
        BF4 & 7.511 & 258 & 113 & 3 & 33.4 & 78 $\pm$ 9 & 2.7 $\pm$ 0.3 & 64 & 822 \\
        BGA & 7.096 & 220 & 138.2 & 3.1 & 36.5 & 131 $\pm$ 20 & 3.3 $\pm$ 0.5 & 121 & 250 \\
        BGB & 7.001 & 174 & 213 & 3.8 & 47.9 & 93 $\pm$ 12 & 2.3 $\pm$ 0.3 & 108 & 479 \\
        \bottomrule
    \end{tabular}
\end{table*}

Across nine $\SI{5}{\milli \metre} \times \SI{5}{\milli \metre}$ test dies, we achieved die-scale relative standard deviations (RSD) in SQUID loop resistances of 2.5\% to 10.3\%, giving a mean RSD of 4.1\%. In comparison with RSDs of approximately 1\% - 4\% achieved for conventional transmon junctions on a wafer-scale \cite{Kreikebaum_2020, Osman_2021,wang2024precision}, these values rather lie at the upper end of the spectrum. However, it should be noted that all dies were experimental in nature, incorporating sweeps of critical fabrication and design parameters such as electron beam exposure doses. For conventional transmon junctions, previous studies have shown that only about 60\% to 70\% of resistance variations can be attributed to fluctuations in junction area, with the remainder largely arising from variations in the junction barrier thickness \cite{Osman_2021}. In contrast, our data suggest that for the significantly larger mergemon junctions, resistance variations are readily explained by area fluctuations with a perfectly coinciding RSD in junction areas of 4.1\%. Assuming the junction current to be carried by a discrete set of conductance channels \cite{Zeng_2015}, this discrepancy could be explained by the fact that larger junctions encompass a broader ensemble of conduction sites, reducing the influence of local barrier inhomogeneities. For a more in-depth description of the fabrication process, see App. \ref{app:fab} and App. Fig. \ref{fig:shadowevaporation}. A breakdown of all room temperature transport measurements can be found in App. Tab. \ref{tab:transport}, as well as App. Fig. \ref{fig:resistancevariation} and \ref{fig:aging}.

\section{Qubit characterization}\label{sec:qubit}

\begin{figure*}
\centering
\subfloat{\includegraphics[width=0.325\textwidth]{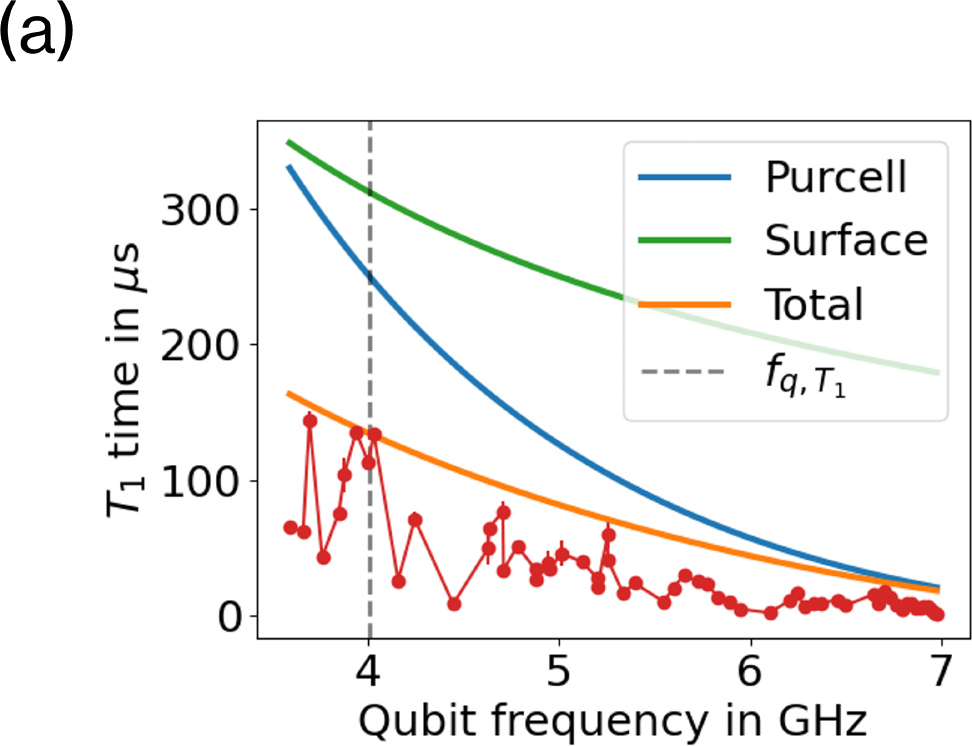}\label{fig:T1a}}
\hfill
\subfloat{\includegraphics[width=0.325\textwidth]{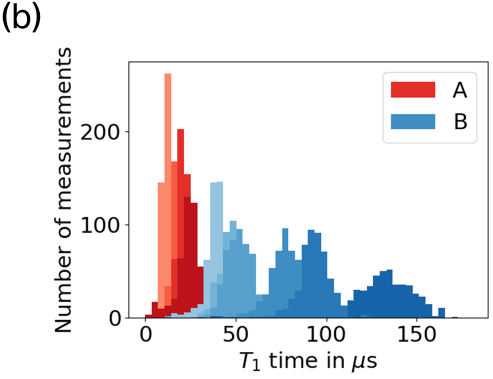}\label{fig:T1b}}
\hfill
\subfloat{\includegraphics[width=0.325\textwidth]{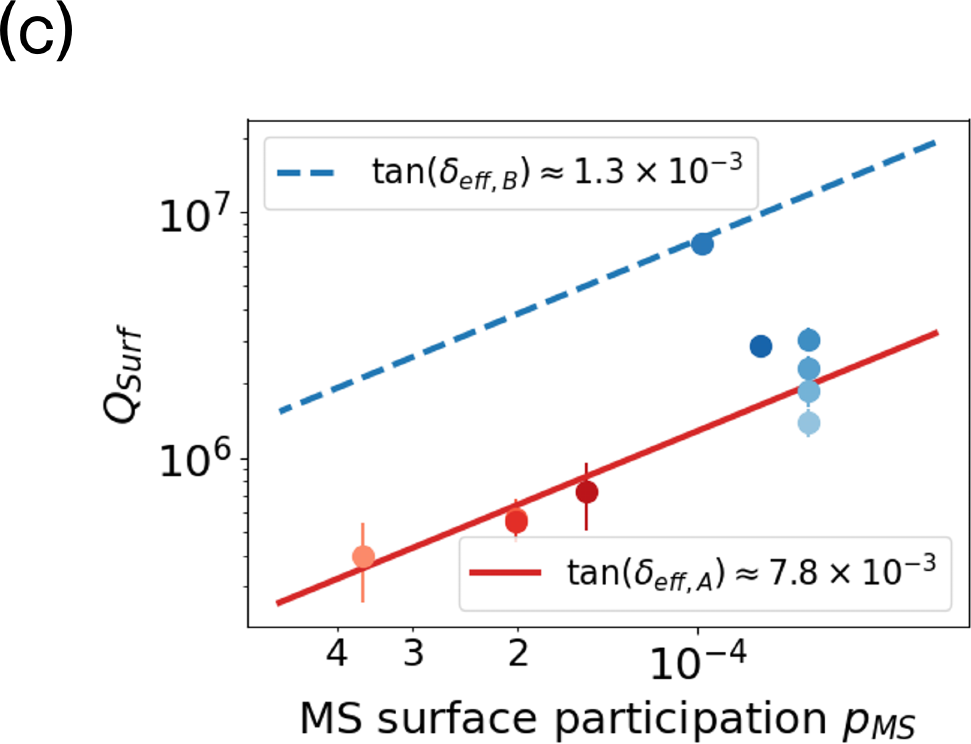}\label{fig:T1c}}
\caption{(a) Measurement of the $T_{1}$ relaxation time of qubit BGA against qubit frequency (red). The $T_{1}$ limitation due to Purcell loss is given by the solid blue line. Further assuming an effective surface loss tangent $\tan(\delta_{\mathrm{eff, B}}) \sim 1.3 \times 10^{-3}$ (solid green line) and a bulk loss tangent of sapphire $\tan(\delta_{\mathrm{S}}) \sim 6.3 \times 10^{-8}$ \cite{Read_2023}, the general trend can be explained by a combination of these loss channels (solid orange line). The vertical, dashed line indicates the frequency at which the long-term $T_{1}$ characterization of qubit BGA presented in (b) was performed. (b) $T_{1}$ time histograms of all mergemon qubits, each recorded over a 12h time period and 500 measurements. Histograms in different shades of red correspond to qubits following design approach A, while shades of blue correspond to approach B. (c) Surface loss limited quality factors $Q_{\mathrm{Surf}}$ of all qubits plotted against the simulated MS interface participation $p_{\mathrm{MS}}$. Data point colors are matched with histogram colors of (b). A fit of the data points of approach A to the participation loss model $Q_{\mathrm{Surf}}^{-1} = p_{\mathrm{MS}} \cdot \tan(\delta_{\mathrm{eff,A}})$ is indicated as a red line. An upper estimate for the surface loss of approach B is indicated by the blue dashed line.}
\label{fig:T1}
\end{figure*}

We investigated a total of ten mergemon qubits inside a dilution refrigerator operated at around $\SI{30}{\milli \kelvin}$ (see App. Fig. \ref{fig:cryogenic} for details on the experimental setup). Table \ref{tab:qubitparameters} summarizes the obtained qubit parameters.

Our measurements support the notion that mergemon qubit transition frequencies are less prone to junction area fluctuations than those of conventional transmons. Despite the large RSDs observed in junction resistances and areas, the measured qubit frequencies at zero flux show a small RSD of 0.98\% between the nominally identical qubits BF1 to BF4 \cite{Kreikebaum_2020, Osman_2021}. Qubits BGA and BGB, despite differing in their respective junction sizes by a factor of approximately 1.3, also display comparable transition frequencies.

The charging energies $E_{\mathrm{C}}$ and Josephson energies $E_{\mathrm{J}}$ were approximated from the $0 \rightarrow 1$ single-photon transition $f_{\mathrm{q}}$ and the $0 \rightarrow 2$ two-photon transition $f_{0 \rightarrow 2}$ using $E_{\mathrm{C}}/h \approx 2(f_{\mathrm{q}}-f_{0 \rightarrow 2})$ and $E_{\mathrm{J}}/h \approx h(f_{\mathrm{q}}+E_{\mathrm{C}}/h)^{2}/8E_{\mathrm{C}}$. With $E_{\mathrm{J}}/E_{\mathrm{C}}$ ratios between 30.6 and 37.4, the mergemon qubits of design approach A reside at the very lower end of the transmon regime, possibly making them suffer from charge dispersion. However, due to the reduced oxidation pressure, qubits following approach B display critical current densities 3 times as high, leading to $E_{\mathrm{J}}/E_{\mathrm{C}}$ ratios between 113 and 213. Recalling the geometric qubit capacitances $C_{\mathrm{geom}}$ extracted from \textit{ANSYS HFSS} simulations (see Tab. \ref{tab:designparameters}), the junction capacitances can be calculated with $C_{\mathrm{JJ}} = (e^2/2E_{\mathrm{C}}-C_{\mathrm{geom}})/2$. Assuming a relative permittivity of 10, the junction barrier thicknesses can be estimated to $\SI{3.1}{\nano \metre}$ for approach A and $\SI{2.7}{\nano \metre}$ for approach B, a substantial increase compared with conventional transmon junctions with 1 to $\SI{2}{\nano \metre}$ \cite{Zeng_2015}.

As exemplified in Fig. \ref{fig:T1a}, we first surveyed the frequency dependence of the $T_{1}$ relaxation time of each qubit using a sparse sampling routine to identify sweet spots in the loss spectrum. Following this survey, a long-term statistical characterization (12h, 500 measurements) was performed at the optimal frequency point $f_{\mathrm{q,T_{1}}}$ (see App. Tab. \ref{tab:purcell}), yielding the histograms presented in Fig. \ref{fig:T1b}. Time-averaged $T_{1}$ values and corresponding quality factors $Q_{\mathrm{q}} = 2\pi f_{\mathrm{q,T_{1}}}T_{1}$ can be found in Tab. \ref{tab:qubitparameters}. An exemplary $T_{1}$ measurement is given in App. Fig. \ref{fig:single_T1}. With some of the qubits being dominated by Purcell loss for frequencies closer to the resonator, and locally by strongly coupled junction TLS (see Purcell loss line and broad dips in Fig. \ref{fig:T1a}), this procedure was necessary to assess the intrinsic $T_{1}$ limitations of the mergemon qubits.

The quality factors of the approach A devices, which feature junction participation ratios consistent with previously reported mergemon geometries (see Tab. \ref{tab:designparameters}), lie between 0.4 and $0.7 \times 10^{6}$. These values align closely with the unannealed devices reported by Mamin et al. (0.33 to $1.1\times 10^6$), while remaining below their postannealing benchmarks (0.9 to $2.2\times 10^6$) \cite{Mamin_2021}. By contrast, the mergemon qubits of approach B, which feature a reduced junction participation, yield substantially higher quality factors in the range of 1.3 to $3.3 \times 10^{6}$. Notably, this range exceeds both the approach A qubits and the unannealed devices in the literature, matching or exceeding even the annealed mergemons reported in Ref. \cite{Mamin_2021}.

We can analyze this trend within the framework of the participation loss model. Here, the measured quality factors $Q_{\mathrm{q}}$ can be expressed in terms of the individual loss channels \cite{Wang_2015}

\begin{equation}\label{eq:loss}
\begin{split}
    Q_{\mathrm{q}}^{-1} = {} & p_{\mathrm{MS}} \tan(\delta_{\mathrm{eff}}) + p_{\mathrm{JJ}} \tan(\delta_{\mathrm{JJ}}) \\
    & + p_{\mathrm{S}} \tan(\delta_{\mathrm{S}}) + Q_{\mathrm{Purcell}}^{-1},
\end{split}
\end{equation}

where $p_{\mathrm{MS}}$, $p_{\mathrm{JJ}}$, and $p_{\mathrm{S}}$ denote the participation ratios for the metal-substrate interface, the Josephson junction, and the bulk substrate, respectively. The specific values for all qubits are provided in Tab. \ref{tab:designparameters}. The effective surface loss tangent is given by $\tan(\delta_{\mathrm{eff}}) = \tan(\delta_{\mathrm{MS}}) + \frac{p_{\mathrm{SA}}}{p_{\mathrm{MS}}}\tan(\delta_{\mathrm{SA}}) + \frac{p_{\mathrm{MA}}}{p_{\mathrm{MS}}}\tan(\delta_{\mathrm{MA}})$ and needs to be defined separately for approach A and B due to differing fabrication techniques and $p_{\mathrm{i}}/p_{\mathrm{MS}}$ ratios (see App. Fig. \ref{fig:surface}). Further, $\tan(\delta_{\mathrm{JJ}})$ is the loss tangent of the Josephson junction and $\tan(\delta_{\mathrm{S}})$ that of the substrate. Finally, $Q_{\mathrm{Purcell}} = 2\pi f_{\mathrm{q,T_{1}}} T_{1}^{\mathrm{Purcell}}$ denotes the limitation due to Purcell loss. A detailed analysis of each device's Purcell limitation can be found in App. \ref{app:purcell}. The extracted qubit-resonator coupling strengths at zero flux $g_{0}/2\pi$ and resulting Purcell limitations $T_{1}^{\mathrm{Purcell}}$ are listed in Tab. \ref{tab:qubitparameters}.

Given a bulk loss tangent of sapphire $\tan(\delta_{\mathrm{S}}) \sim 6.3 \times 10^{-8}$ \cite{Read_2023}, we find the quality factor limitation due to dielectric loss to the substrate for all mergemon qubits to lie between $Q_{\mathrm{S}} \sim 1$ and $5 \times 10^{8}$. Consequently, this loss channel is 2 orders of magnitude too weak to account for the performance variations observed between the designs of approach A and B. Notably, this negligible substrate sensitivity highlights a key advantage of the mergemon architecture, as bulk substrate losses present a much more severe coherence bottleneck for conventional transmons of approximately $2 \times 10^{7}$ for our baseline transmon and for the value obtained in Ref. \cite{Read_2023}.

Although Purcell loss contributes significantly to the relaxation of certain devices (e.g., $T_{\mathrm{1, BGA}} \approx 0.5$ $T_{\mathrm{1, BGA}}^{\mathrm{Purcell}}$), no correlation is found between the calculated Purcell limits and the observed trends in quality factors across the different designs.

While junction and surface loss are treated analogously in Eq. \ref{eq:loss}, they are rooted in fundamentally different defect landscapes. Whereas surface loss originates from a large-volume, weak-field environment, leading to relaxation into a large ensemble of weakly coupled surface TLS ($N \sim 10^{2}$ \cite{Cho_2023}), junction loss originates from a comparably small-volume, high-field environment, leading to a small set of strongly coupled junction TLS governing the loss channel ($N \sim 10^{0}$ \cite{Lisenfeld_2019}). To investigate whether the qubits are dominated by surface- or junction-loss, we first hypothetically assume that the observed maximum qubit quality factors correspond to a case where our measurement routine successfully isolated optimal operating points between these localized junction TLS resonances and hence junction-loss was negligible. The qubit quality factor due to surface loss then reads

\begin{equation}
    Q_{\mathrm{Surf}} \approx \left(Q_{\mathrm{q}}^{-1} - p_{\mathrm{S}} \tan(\delta_{\mathrm{S}}) - Q_{\mathrm{Purcell}}^{-1}\right)^{-1},
\end{equation}

such that $Q_{\mathrm{Surf}}^{-1} = p_{\mathrm{MS}} \tan(\delta_{\mathrm{eff}})$. The resulting values for all mergemon qubits are plotted in Fig. \ref{fig:T1c} and observed to overall increase with decreasing $p_{\mathrm{MS}}$. Qubits of approach A closely follow this trend, indicating that surface loss is indeed the dominant source of energy relaxation. A fit to the participation loss model (see red line in Fig. \ref{fig:T1c}) gives a relatively high surface loss tangent of $\tan(\delta_{\mathrm{eff,A}}) = 7.8 \times 10^{-3}$ for approach A compared with values obtained for conventional aluminum-on-sapphire transmon qubits and CPW resonators of $\tan(\delta_{\mathrm{eff}}) \sim$ 1 to $6 \times 10^{-3}$ \cite{Wang_2015, hedrick2026quantifyingsurfacelossessuperconducting}. However, this discrepancy can be explained by the physical realization of the qubits. Unlike a conventional transmon, the entire qubits needed to be fabricated by additive lift-off on a pre-etched substrate, both of which are known to cause increased surface loss \cite{weeden2025statisticsstronglycoupleddefects, Wang_2015, lisenfeld2026mapping, smirnov2023wiringsurfacelosssuperconducting}. On the other hand, the $Q_{\mathrm{Surf}}$ values of qubits following approach B seem to increase with $p_{\mathrm{MS}}$ and show a large spread among identically designed devices, indicating that surface loss is not the dominant source of energy relaxation. However, with all of the qubits of approach B being fabricated identically in the same evaporation run, we can get an upper estimate for the surface loss tangent of approach B by only fitting to the highest-performing qubit BGA. As shown in Fig. \ref{fig:T1a}, a reasonable loss tangent of $\tan(\delta_{\mathrm{eff,B}}) \approx 1.3 \times 10^{-3}$ neatly fits the whole $T_{1}$ vs frequency spectrum of the device. Therefore, the enlargement of the qubit islands not only achieved surface participation ratios much lower than those of approach A and conventional transmons alike, but also allows for a cleaner subtractive fabrication technique, further minimizing the surface loss tangent. As a result, surface loss is suppressed to the point where the device is limited by junction loss, as intended by the mergemon paradigm.

Hence, instead of successfully probing the qubits within a window where surface loss dominates, the $T_{1}$ characterizations for approach B were likely performed on the overlapping spectral tails of a few strongly coupled junction TLS. This claim is backed up by the large device-to-device fluctuations in $Q_{\mathrm{Surf}}$ observed among the identically designed qubits BF1-4, which can be readily explained by the small-sample variance in this framework. Assuming that now surface loss is negligible, we can apply an analogous treatment as above, defining the junction loss limited quality factor $Q_{\mathrm{JJ}} \equiv Q_{\mathrm{Surf}}$ to extract the junction loss tangent (see App. \ref{app:junction}). In comparison to typical literature values for amorphous aluminum oxide $\tan(\delta_{\mathrm{AlOx}}) \sim 10^{-3}-10^{-2}$ \cite{hedrick2026quantifyingsurfacelossessuperconducting, Martinis2}, we find a relatively low value of $\tan(\delta_{\mathrm{JJ,B}}) \approx 6 \times 10^{-6}$. However, reports in the literature occasionally extract even lower baseline values for the junction oxide's loss tangent from participation loss considerations in the order of $\tan(\delta_{\mathrm{JJ}}) \sim 1 \times 10^{-7}$\cite{Mamin_2021, hedrick2026quantifyingsurfacelossessuperconducting}. This discrepancy is rooted in a breakdown of the simple participation loss model for such small volumes and consequently diminishing numbers of TLS. Just as including the low-volume, high-field near-junction region in Ref. \cite{Wang_2015} resulted in an unphysical, negative offset, we also needed to include a negative offset $C \approx -4.8 \times 10^{-6}$ to correctly fit the data. Therefore, while this empirical approach serves as a useful baseline comparison, it motivates the more rigorous, microscopic TLS-based analysis of the junction oxide's loss tangent presented in the next section.

We refrain from a systematic analysis of mergemon dephasing times, because our experimental setup is tailored to provide means for TLS tuning by applied mechanical strain and dc-electric fields, for which a piezo actuator and gate electrodes are installed near the qubit sample which introduce additional low-frequency noise that dominates qubit dephasing.

\section{TLS spectroscopy}\label{sec:tls}

\begin{figure*}[tbh!]
\centering
\subfloat{\includegraphics[width=0.325\textwidth]{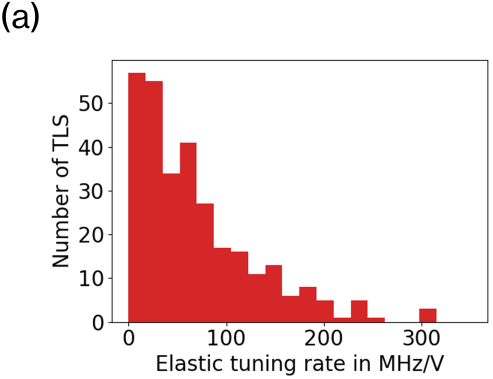}\label{fig:TLSa}}
\hfill
\subfloat{\includegraphics[width=0.325\textwidth]{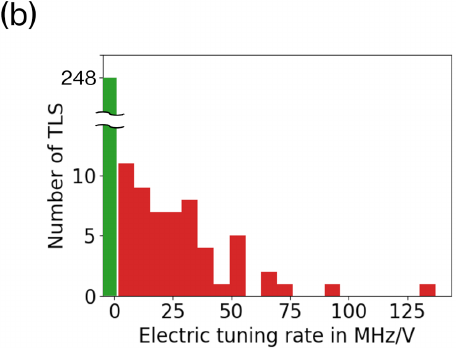}\label{fig:TLSb}}
\hfill
\subfloat{\includegraphics[width=0.325\textwidth]{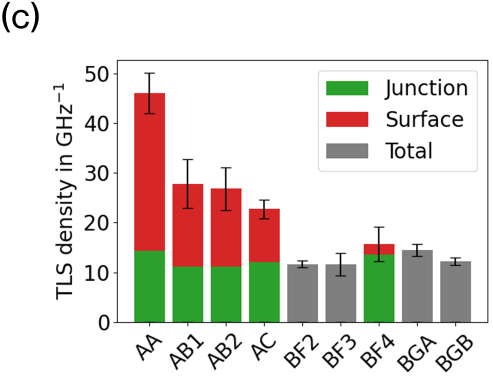}\label{fig:TLSc}}
\vskip 1em
\centering
\subfloat{\includegraphics[width=0.325\textwidth]{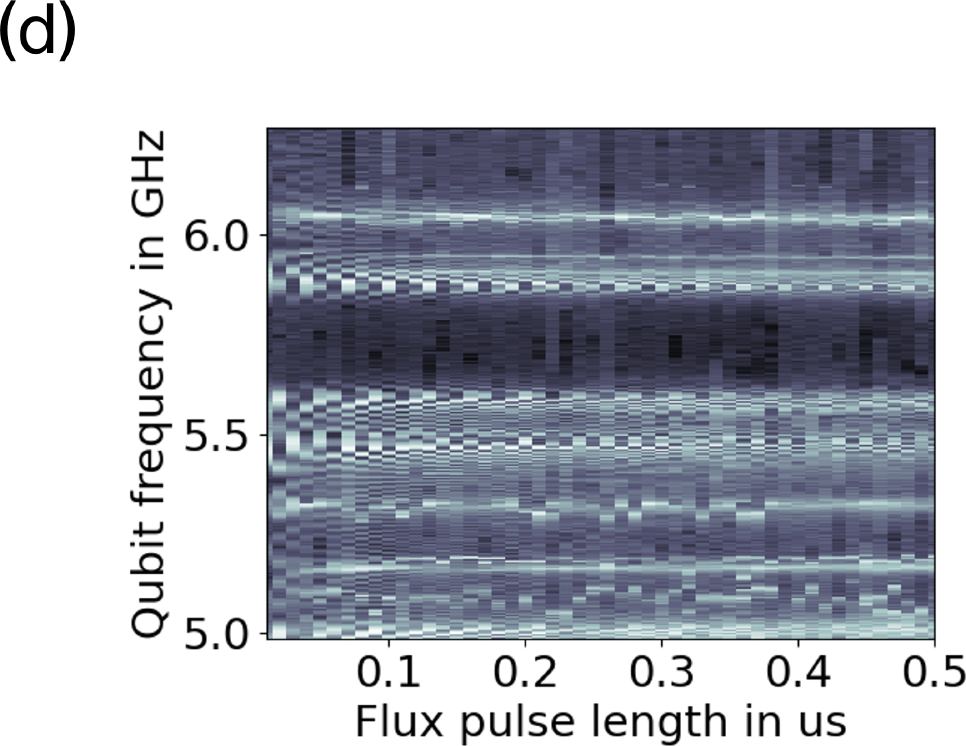}}\label{fig:TLSd}
\subfloat{\includegraphics[width=0.325\textwidth]{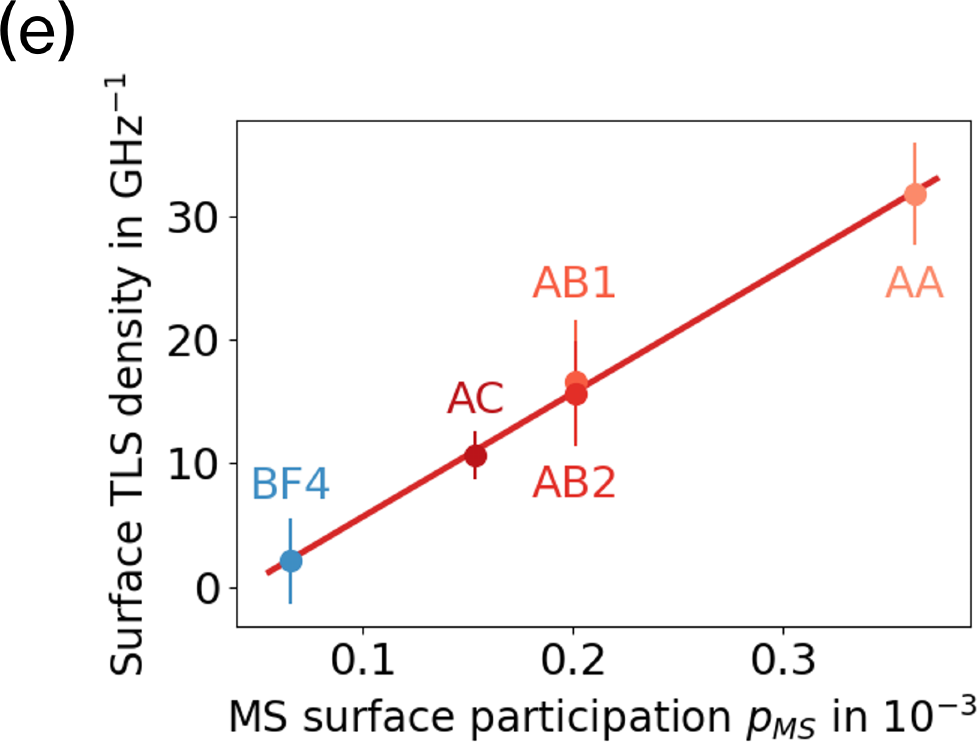}\label{fig:TLSe}}
\hfill
\caption{(a) and (b) Histograms of the elastic and electric TLS tuning strengths, extracted from strain- and electric-field-dependent swap spectroscopy. TLS for which no tuning with electric field could be identified within the fit errors (green in (b)) are classified as junction TLS. (c) Spectral junction TLS density (green) and surface TLS density (red) of the inspected qubits. For qubits in which zero TLS were found that responded to the applied E-field, only the total observed TLS densities (gray) are given since we cannot neglect the small possibility that the gate electrode was not functional.  (d) Swap oscillations recorded between qubit BGB and seven junction TLS to directly extract the respective coupling strengths. (e) Spectral surface TLS densities of the five qubits, from which values could be extracted (see (c)), plotted against the simulated MS surface participation together with a linear fit (red line).}
\label{fig:TLS}
\end{figure*}

To distinguish qubit energy relaxation due to surface- and junction TLS, we performed strain- and electric-field-dependent swap spectroscopy \cite{Lisenfeld_2019}. Therefore, the qubit was prepared in its excited state and tuned to various probe frequencies for a duration of $\SI{10}{\micro \second}$. The remaining qubit population was then measured to estimate the qubit's $T_{1}$ time, which shows minima when the qubit is in resonance with a strongly coupled TLS. In addition, TLS were tuned with dc-electric fields generated by a gate electrode and with strain exerted by a piezo actuator beneath the sample. These fields were swept in an alternating fashion to extract the elastic tuning strength $\gamma_{\mathrm{S}}$ and electric tuning strength $\gamma_{\mathrm{E}}$ of each visible TLS by fitting its resonance frequency to the standard tunneling model

\begin{equation}\label{eq:stm}
    f_{\mathrm{TLS}} = \sqrt{\Delta^{2}+\left(\epsilon_{0}+\gamma_{\mathrm{S}}V_{\mathrm{Piezo}}+\gamma_{\mathrm{E}}V_{\mathrm{dc}}\right)^{2}}.
\end{equation}

Here, $\Delta$ is the tunneling rate, $\epsilon_{0}$ the offset asymmetry energy, $V_{\mathrm{Piezo}}$ the voltage applied to the piezo actuator, and $V_{\mathrm{dc}}$ the voltage applied to the top gate. An exemplary measurement can be found in App. Fig. \ref{fig:tlsspec}.

Thereby, we were able to extract the tuning strengths of a total of 307 TLS over nine mergemon qubits. As shown in Fig. \ref{fig:TLSa} and \ref{fig:TLSb}, both tuning strengths are randomly distributed, except for a large fraction of TLS showing no tuning with electric field within the fit errors. We can use this to classify the observed TLS into junction and surface TLS, since surface TLS are expected to respond to both applied electric and strain fields, while junction TLS are shielded from electric fields and therefore tune only with strain. 

Further normalizing the number of TLS counted per voltage step to the frequency bandwidth of the scan and averaging over all applied tuning values, we extracted the spectral TLS densities of all qubits and TLS types. A comprehensive overview can be found in Fig. \ref{fig:TLSc}. It shall be noted that no TLS tuning with electric field could be observed for qubits BF2, BF3, BGA, and BGB. Invoking the low surface TLS density of qubit BF4, this observation could be explained by a diminishing surface TLS contribution to energy relaxation ($g \ll 1/T_{1}$) among qubits following approach B. On the other hand, it could also indicate a malfunction of the top gate for these measurements. Because of this ambiguity, only the total TLS density is provided in Fig. \ref{fig:TLSc}.

Despite the change in oxidation pressure between approach A and B, all qubits show a similar junction TLS density normalized to the total junction area of about $\SI{6.2}{\per \giga \hertz \per \micro \metre \squared}$. This at first glance unexpected behavior can be explained by the only minor change of approximately 1.15 in barrier thickness, leading to an increase in junction barrier volume, and therefore in the number of junction TLS, but also a decrease in junction field, and therefore in the overall coupling strength and hence detectability \cite{lisenfeld2026mapping}. Recalling the estimated junction barrier thicknesses of 2.7 to $\SI{3.1}{\nano \metre}$ (see Sec. \ref{sec:qubit}), this corresponds to a volume density between 2000 and $\SI{2300}{\per \giga \hertz \per \micro \metre \cubed}$. In comparison to densities of 200 to $\SI{1800}{\per \giga \hertz \per \micro \metre \cubed}$ reported in previous studies for AlOx \cite{Lisenfeld_2019, Mueller_2019, Bilmes_2020, PhysRevB.77.180502, Bilmes_2021}, these values are quite high. However, the derived densities strongly depend on the used method (swap spectroscopy vs qubit spectroscopy), qubit coherence (hundreds of nanoseconds in phase qubits vs tens of microseconds in transmons), and electric field strength inside the inspected junction (stray junction vs tunnel junction). With long $T_{1}$ relaxation times and strong junction fields, leading to large coupling strengths, our measurements are very sensitive to junction TLS, potentially explaining these high values. In addition, a genuinely higher defect density may also arise from the thicker junction barrier, leading to an increased dangling bond density \cite{Molina_Ruiz_2021, Bilmes_2021}, and the higher complexity in junction fabrication, providing more room for the incorporation of TLS. As shown in Fig. \ref{fig:TLSd}, we recorded coherent swap oscillations between qubit BGB and seven junction TLS to directly extract the respective coupling strengths $g/2\pi$. These range from 3.9 to $\SI{24.2}{\mega \hertz}$, corresponding to electric dipole moments of 0.1 to 0.7e$ \si{\angstrom}$ parallel to the junction field of about $\SI{1.5}{\kilo \volt \per \metre}$. From the average of the calculated dipole moments and the volume junction TLS density of $\SI{2300}{\per \giga \hertz \per \micro \metre \cubed}$, we can estimate the loss tangent of the barrier oxide \cite{Bilmes_2021} to approximately $\tan(\delta_{\mathrm{JJ}}) \approx 1.2 \times 10^{-3}$. This value is in good agreement with previously reported results for the junction oxide \cite{Martinis2}, and the loss tangent we found for the surface oxides of approach B. All calculations can be found in App. \ref{app:TLS}. A breakdown of the TLS parameters obtained from swap oscillations can be found in App. Tab. \ref{tab:swap}.

The spectral surface TLS densities of mergemon qubits following design approach A range from 10.7 to $\SI{31.8}{\per \giga \hertz}$. Hence, in comparison with results obtained for conventional transmon qubits of approximately $\SI{25}{\per \giga \hertz}$ \cite{Lisenfeld_2019, Bilmes_2020}, we find a similar surface TLS density. With corresponding shares of 47\% to 69\% of the total TLS density, surface loss still represents a significant contribution to decoherence in these qubits, despite their drastically reduced footprint and increased junction participation. On the other hand, for qubit BF4, we obtain a 10 times lower surface TLS density of only $\SI{2.1}{\per \giga \hertz}$, accounting for 13\% of the total TLS density. This indicates that surface loss is significantly reduced in qubits following design approach B, leading to the expected limitation due to junction loss. Invoking the linear dependency of surface TLS density on $p_{\mathrm{MS}}$ plotted in Fig. \ref{fig:TLSe}, this reduction can be again attributed to the lower $p_{\mathrm{MS}}$ of type B qubits. Therefore, careful design considerations aimed at minimizing the surface participation ratio remain essential, even for the mergemon approach.

\section{Conclusion and Outlook}\label{sec:con}

We demonstrated flux-tunable mergemon qubits achieving mean $T_{1}$ relaxation times of up to $\SI{130}{\micro \second}$, corresponding to quality factors of up to $3.3 \times 10^{6}$. Thus, compared with previously reported results for mergemon qubits, our devices exhibit quality factors that meet or exceed the current state of the art and are on par with those of conventional transmon qubits fabricated using similar technology.

Utilizing strain- and electric-field-dependent swap spectroscopy, we showed that careful design considerations are essential for mergemon qubits to avoid $T_{1}$ limitations due to surface loss. While non-optimally designed devices exhibited surface TLS densities comparable to those of transmons, geometry optimization led to a tenfold reduction.

Furthermore, we developed a fabrication technique, capable of realizing the needed, thicker-than-usual Josephson junction barriers without relying on hours long oxidations at extreme oxygen pressures. Our findings indicate that, in contrast to conventional transmon junctions, the observed resistance fluctuations ($\sigma_{\mathrm{R}} \approx 4.1\%$) can be entirely attributed to junction area fluctuations ($\sigma_{\mathrm{A}} \approx 4.1\%$) - a potentially easier to stabilize parameter than barrier thickness variations. Additionally, we find that mergemon qubit transition frequencies are less prone to junction area fluctuations, with a relative standard deviation of only $\sigma_{\mathrm{f}} \approx 0.98\%$, despite the large fluctuations in junction area. Given the need for precise transition frequency control in large-scale quantum processors, this appears as another advantage of the mergemon approach.

It shall also be noted that mergemon qubits are an excellent testbed for the mitigation of junction TLS. With a long energy relaxation time and strong, well defined electric field inside the junction, we were able to resolve a high spectral density of junction TLS and determine their coupling strengths as well as dipole moments. Future research could aim at utilizing these properties to perform extensive studies of junction TLS, potentially providing a deeper insight into potential mitigation strategies.

Finally, due to the large junction participation and small footprint of the mergemon qubit, surface and substrate losses are expected to play a less pronounced role at high coherence compared with conventional transmon qubits \cite{charpentier2025universalscalingmicrowavedissipation}. Thus, instead of optimizing large volumes of dielectric, here only the comparatively small junction barrier needs to be cleaned from TLS. While postprocessing techniques like thermal and alternating-bias assisted annealing already indicate some degree of improvements in junction performance \cite{Mamin_2021, Pappas_2024}, further breakthroughs in fabrication could quickly lead to mergemon qubits outperforming transmons.

\section*{Acknowledgments}

We thank Alexander Bilmes for fruitful discussions, as well as for his contributions to the experimental setup and qubit fabrication. We acknowledge the contributions of Hannes Rotzinger to the experimental setup. We are grateful for the clean room facilities provided for the fabrication by the Nanostructure Service Laboratory (NSL) at KIT. We thank Lucas Radtke and Silvia Diewald for their contributions to the fabrication. This study was funded by Google.\\
\newline
E.D. designed the B type samples, fabricated the samples, performed the measurements, analyzed the data, and wrote the manuscript. B.B. designed the A type mergemon qubits and contributed to sample fabrication, measurement, and writing of the manuscript. S.D. designed the $\lambda/2$ CPW resonators employed in A type samples. A.V.U. provided the experimental means and contributed to the writing of the manuscript. J.L. devised the experiment, supervised the project, implemented the experimental setup, and contributed to the writing of the manuscript.

\section*{Data availability}

The data that support the findings of this article are not publicly available upon publication because it is not technically feasible and/or the cost of preparing, depositing, and hosting the data would be prohibitive within the terms of this research project. The data are available from the authors upon reasonable request.

\appendix

\section{ANSYS HFSS simulations}\label{app:simulation}

The electric field distributions, geometric capacitances $C_{\mathrm{geom}}$ and energy participation ratios $p_{\mathrm{i}}$ of all six mergemon qubit designs, as well as one of our standard transmon designs, were simulated using \textit{ANSYS HFSS}'s Eigenmode Solver. In all mergemon models, the simulation volume was restricted to a $\SI{500}{\micro \metre} \times \SI{500}{\micro \metre} \times \SI{1500}{\micro \metre}$ region centered around the qubit. All thin-film structures, including qubit islands, Josephson junctions, resonators, bias lines, and ground planes, were modeled as 2D sheets on a $\SI{500}{\micro \metre} \times \SI{500}{\micro \metre} \times \SI{500}{\micro \metre}$ sapphire substrate. Except for the junctions, all sheets were assigned the ``Perfect E'' boundary condition, while the junction sheets were assigned the ``Lump RLC'' boundary condition. The applied junction capacitances $C_{\mathrm{JJ}}$ were estimated from the measured qubit charging energies (see Sec. \ref{sec:qubit} of the main text) and simulated geometric capacitances $C_{\mathrm{geom}}$ to approximately $\SI{28.5}{\femto \farad}$ for the $\SI{1}{\micro \metre \squared}$ junctions of approach A, $\SI{34}{\femto \farad}$ for the $\SI{1}{\micro \metre \squared}$ junctions of approach B, $\SI{36.5}{\femto \farad}$ for $\SI{1.25}{\micro \metre \squared}$, and $\SI{48}{\femto \farad}$ for $\SI{1.5}{\micro \metre \squared}$. For straightforward comparison, the junction inductances $L_{\mathrm{JJ}}$ were adjusted such that the simulated eigenfrequency of each qubit was approximately $f_{\mathrm{q}} \approx \SI{5}{\giga \hertz}$. Along the edges of qubit sheets and the entire SQUID loop, the precision of the adaptive mesh solver was set to a minimum of $\SI{100}{\nano \metre}$. The respective geometric capacitances were obtained from the set eigenfrequencies, junction capacitances, and junction inductances

\begin{equation}
C_{\mathrm{geom}} = \frac{2}{(2\pi f)^2L_{\mathrm{JJ}}} - 2C_{\mathrm{JJ}}.
\end{equation}

The combined electric energy participation ratio of both Josephson junctions is then readily obtained from the respective capacitance ratio

\begin{equation}
p_{\mathrm{JJ}} = \frac{2C_\mathrm{JJ}}{2C_\mathrm{JJ}+C_{\mathrm{geom}}}.
\end{equation}

In order to obtain the surface energy participation ratios and surface electric field plots, the electric field energy outside of the lumped-element Josephson junctions is then normalized to

\begin{equation}
E_{\mathrm{ext}} = \frac{E_{\mathrm{tot}}C_\mathrm{geom}}{2C_\mathrm{JJ}+C_{\mathrm{geom}}},
\end{equation}

where $E_{\mathrm{tot}} = hf_{\mathrm{q}}/2$. The energy stored in the thin, lossy dielectric layers at the metal-substrate (MS), substrate-air (SA) and metal-air (MA) interfaces is then estimated from the surface integrals \cite{Wenner_2011}

\begin{align}
E_{\mathrm{MS}} &= \frac{1}{4}\epsilon_{0} t \epsilon_{\mathrm{r}} \int_{MS} |E_{s,\perp}|^2 dA, \\
E_{\mathrm{MA}} &= \frac{1}{4}\epsilon_{0} t \epsilon_{\mathrm{r}}^{-1} \int_{MA} |E_{a,\perp}|^2 dA, \\
E_{\mathrm{SA}} &= \frac{1}{4}\epsilon_{0} t \left[\epsilon_{\mathrm{r}}^{-1}\int_{SA} |E_{a,\perp}|^2 dA + \epsilon_{\mathrm{r}} \int_{SA} |E_{a,\parallel}|^2 dA \right].
\end{align}

Here, $E_{\mathrm{a}}$ ($E_{\mathrm{s}}$) is the peak electric field in the air (substrate) outside the interface and $E_{\parallel}$ ($E_{\perp}$) is the electric field component parallel (perpendicular) to the interface. The dielectric layers are assumed to have a relative permittivity $\epsilon_{\mathrm{r}} = 10$ and thickness $t = \SI{3}{\nano \metre}$. Finally, the energy stored in the substrate is determined by

\begin{equation}
E_{\mathrm{Sub}} = \frac{1}{4}\epsilon_{0} \epsilon_{\mathrm{r}} \int_{V} |E_{s}|^2 dV,
\end{equation}

where $\epsilon_{\mathrm{r}} = 10$ is assumed for sapphire and the volume V spans the whole substrate. The respective energy participation ratios are then defined by

\begin{equation}
p_{\mathrm{i}} = \frac{E_{\mathrm{i}}}{E_{\mathrm{tot}}}.
\end{equation}

\begin{figure*}[tbh!]
\centering
\subfloat{\includegraphics[width=0.325\textwidth]{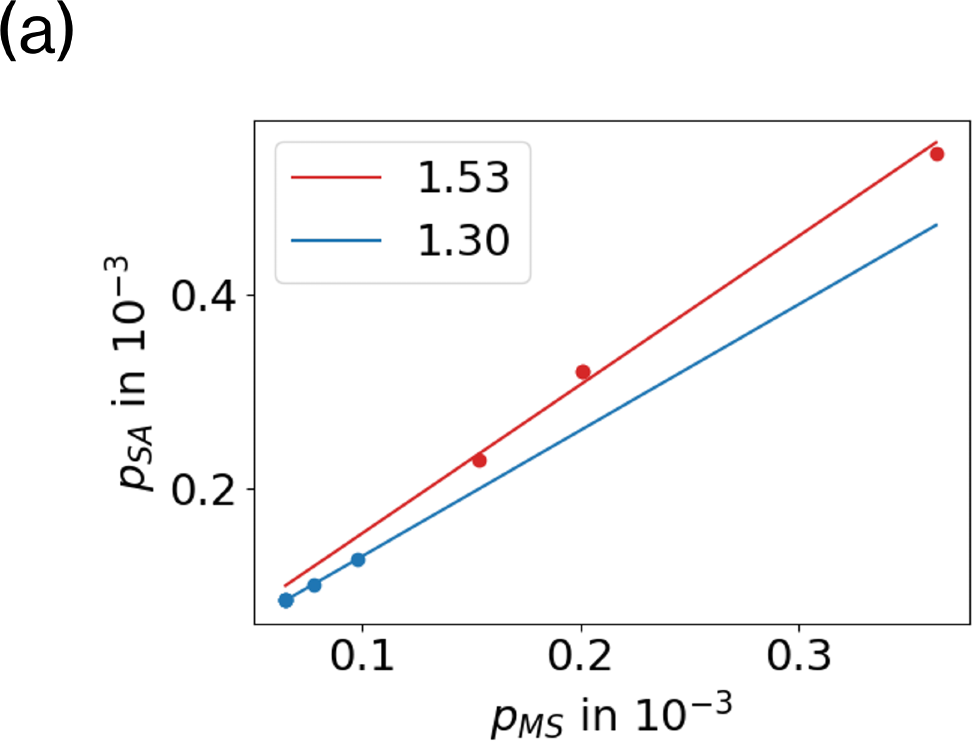}\label{fig:surfacea}}
\hspace{1cm}
\subfloat{\includegraphics[width=0.325\textwidth]{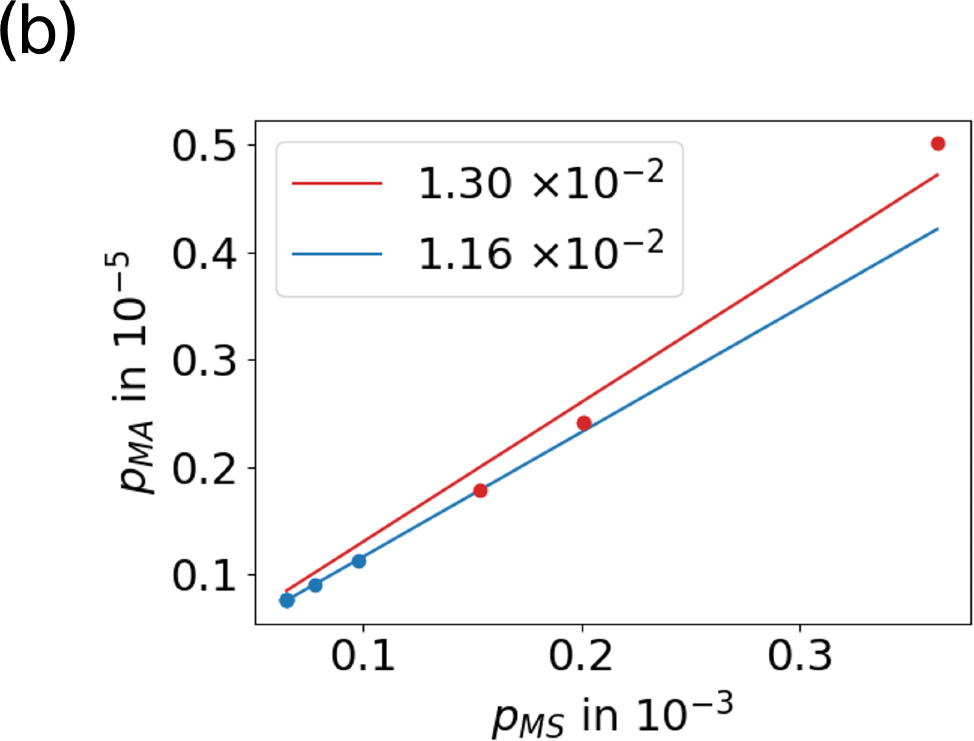}\label{fig:surfaceb}}
\caption{SA participation (a) and MA participation (b) plotted against MS participation for all mergemon qubit designs. Ratios $p_{\mathrm{SA}}/p_{\mathrm{MS}}$ and $p_{\mathrm{MA}}/p_{\mathrm{MS}}$ are extracted separately for design approach A (red) and B (blue) from linear fits indicated by the respective solid lines.}
\label{fig:surface}
\end{figure*}

The obtained surface participation ratios $p_{\mathrm{SA}}$ and $p_{\mathrm{MA}}$ are plotted against the MS surface participation $p_{\mathrm{MS}}$ in Fig. \ref{fig:surface}. The resulting $T_{1}$ limitation due to dielectric loss to these interface layers is given by

\begin{equation}
\frac{1}{T_{\mathrm{1,Surf}}} = \omega_q \sum_{i} p_{\mathrm{i}} \tan(\delta_{\mathrm{i}}).
\end{equation}

With all three participation ratios changing approximately in proportion among mergemons following the same design approach (see linear fits in Fig. \ref{fig:surface}), we can define an effective loss tangent normalized to the MS participation

\begin{equation}
\tan(\delta_{\mathrm{eff}}) = \tan(\delta_{\mathrm{MS}})+ \frac{p_{\mathrm{SA}}}{p_{\mathrm{MS}}}\tan(\delta_{\mathrm{SA}})+ \frac{p_{\mathrm{MA}}}{p_{\mathrm{MS}}}\tan(\delta_{\mathrm{MA}}),
\end{equation}

such that $Q_{\mathrm{Surf}}^{-1} = p_{\mathrm{MS}} \tan(\delta_{\mathrm{eff}})$. The ratios $p_{\mathrm{SA}}/p_{\mathrm{MS}}$ and $p_{\mathrm{MA}}/p_{\mathrm{MS}}$ are extracted separately for design approach A and B from linear fits as shown in Fig. \ref{fig:surface}.

\section{Fabrication}\label{app:fab}

\begin{figure*}[tbh!]
\centering
\subfloat[Deposition of \SI{30}{\nano \metre} aluminum at \SI{1}{\nano \metre \per \second} and an angle of \ang{59}, forming the bottom electrode.]{\includegraphics[width = 0.33\textwidth]{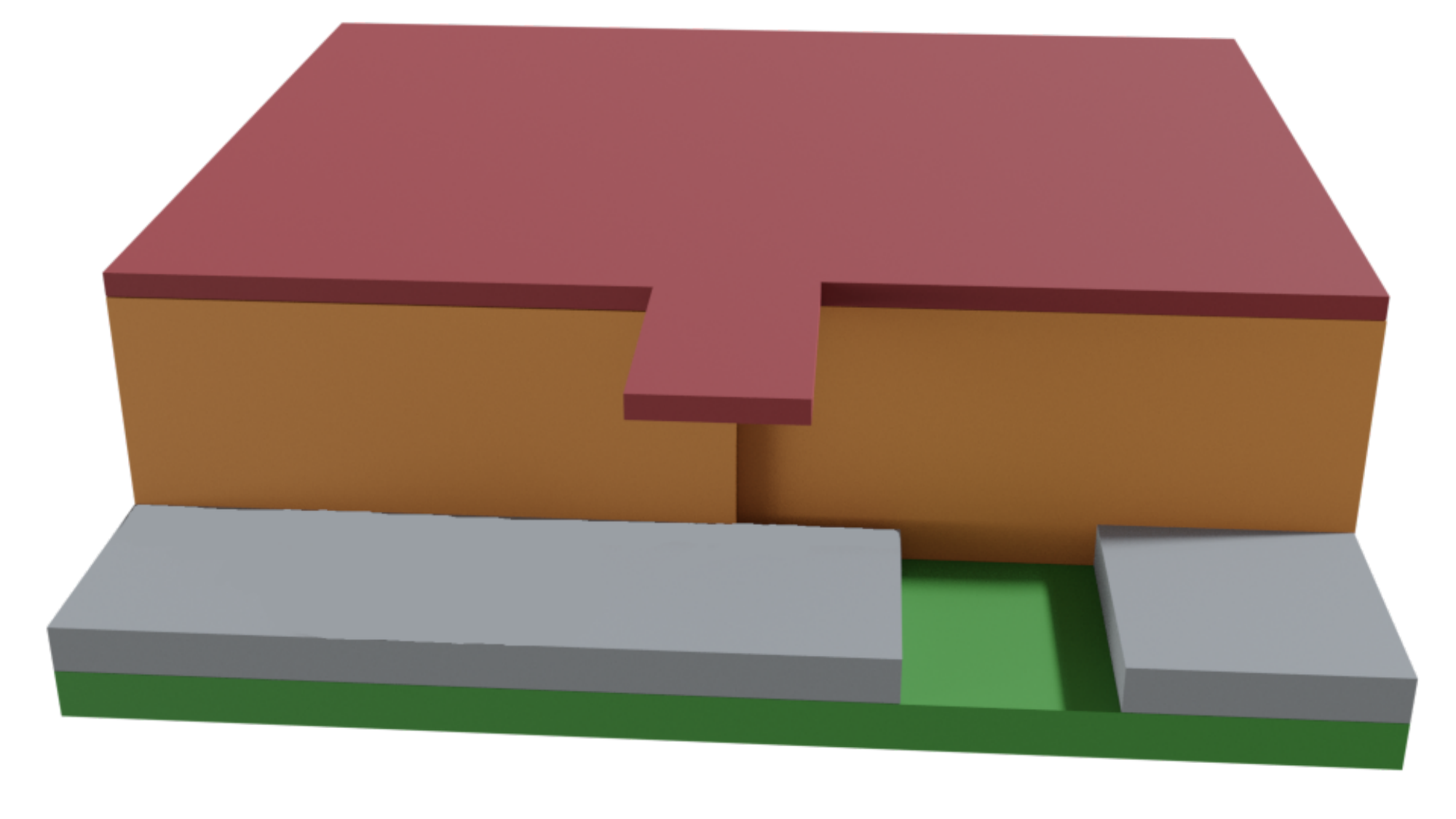}}
\hfill
\subfloat[Oxidation at \SI{130}{\milli \bar} (A) or \SI{25}{\milli \bar} (B) for $\SI{20}{\minute}$, forming the first part of the junction barrier.]{\includegraphics[width = 0.33\textwidth]{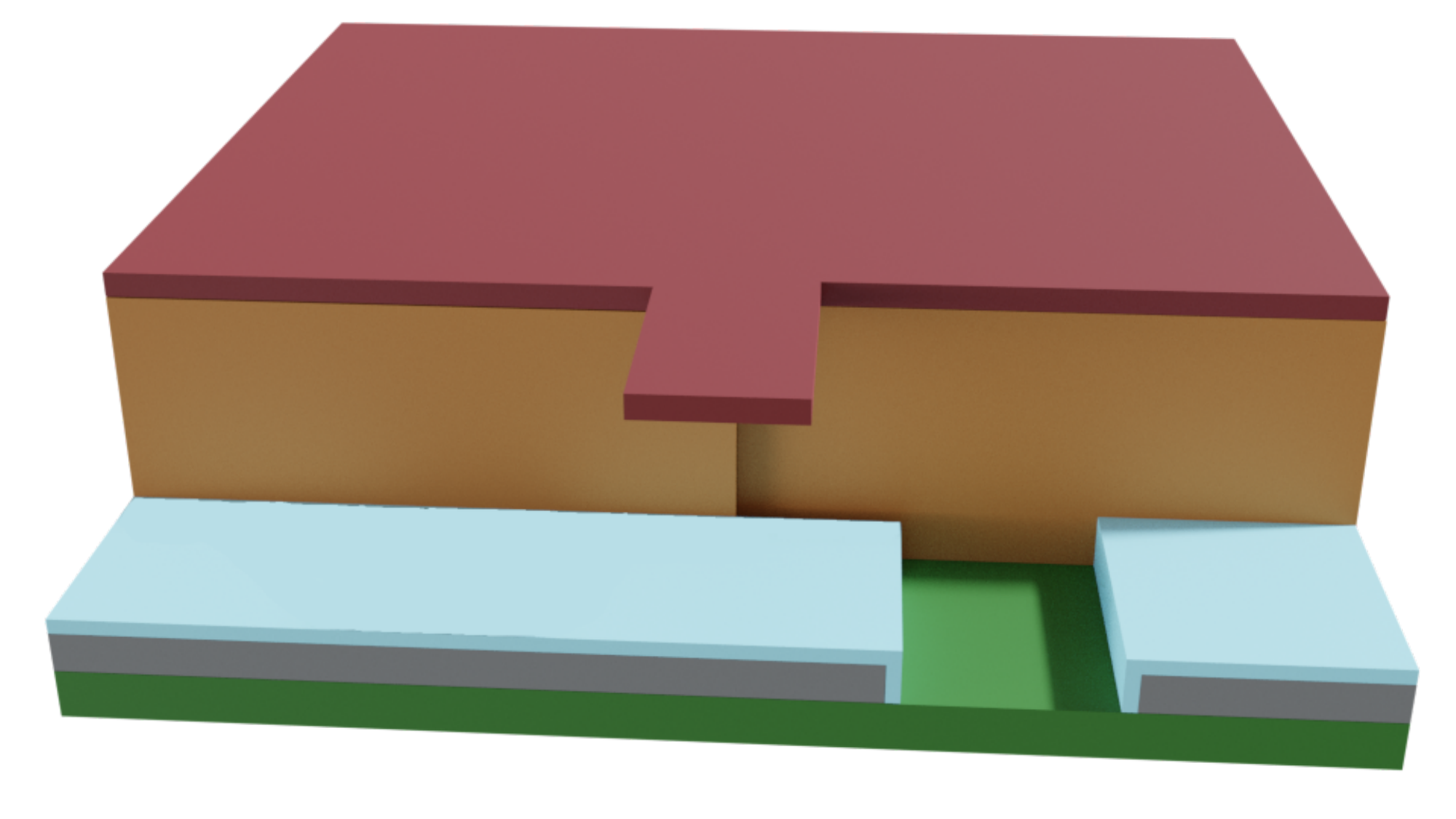}}
\hfill
\subfloat[Deposition of $\SI{0.1}{\nano \metre}$ aluminum at \SI{0.1}{\nano \metre \per \second} and an angle of \ang{59}.]{\includegraphics[width = 0.33\textwidth]{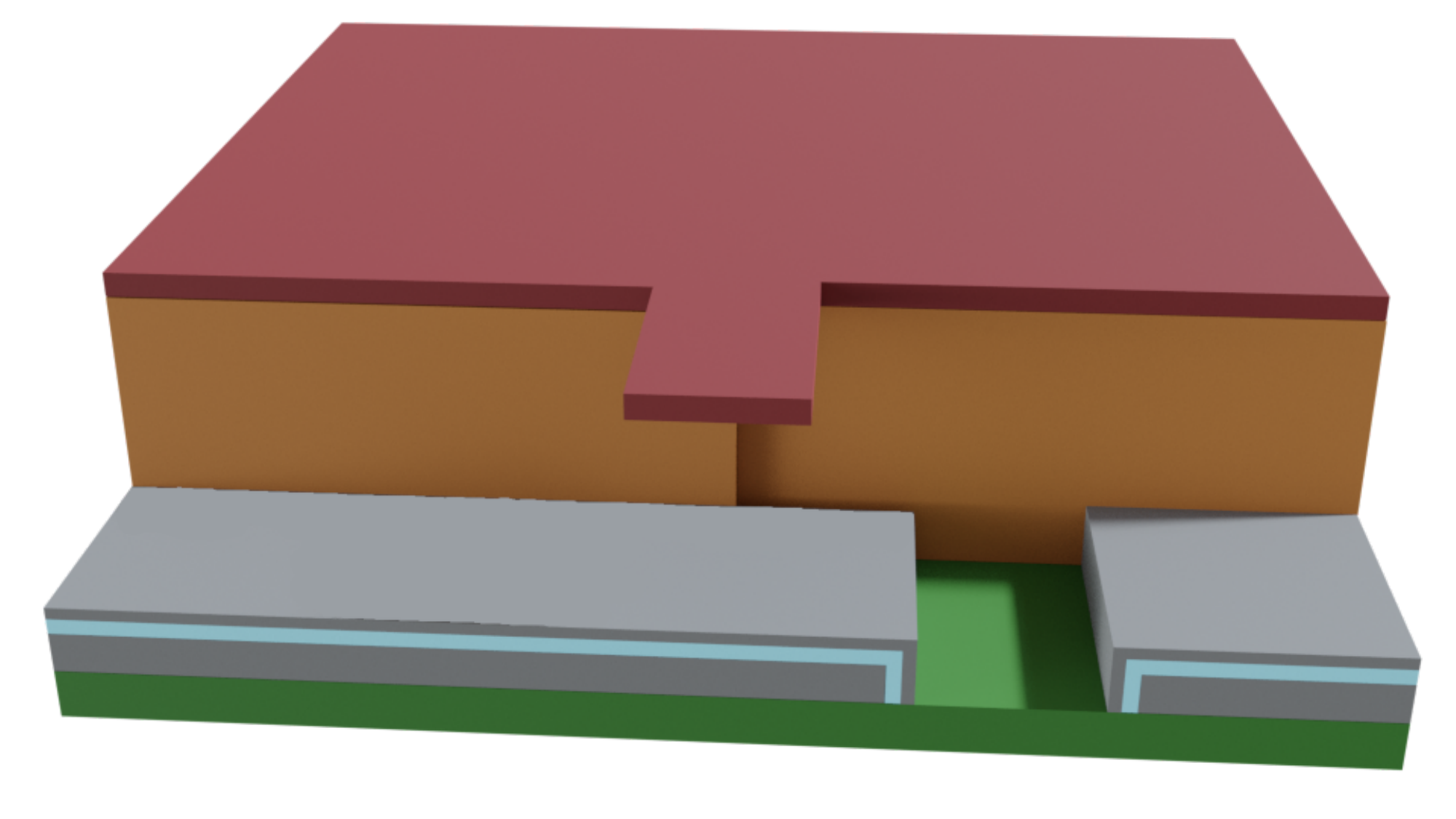}}
\vskip 1em
\subfloat[Oxidation at \SI{130}{\milli \bar} (A) or \SI{25}{\milli \bar} (B) for $\SI{20}{\minute}$, completing the junction barrier.]{\includegraphics[width = 0.33\textwidth]{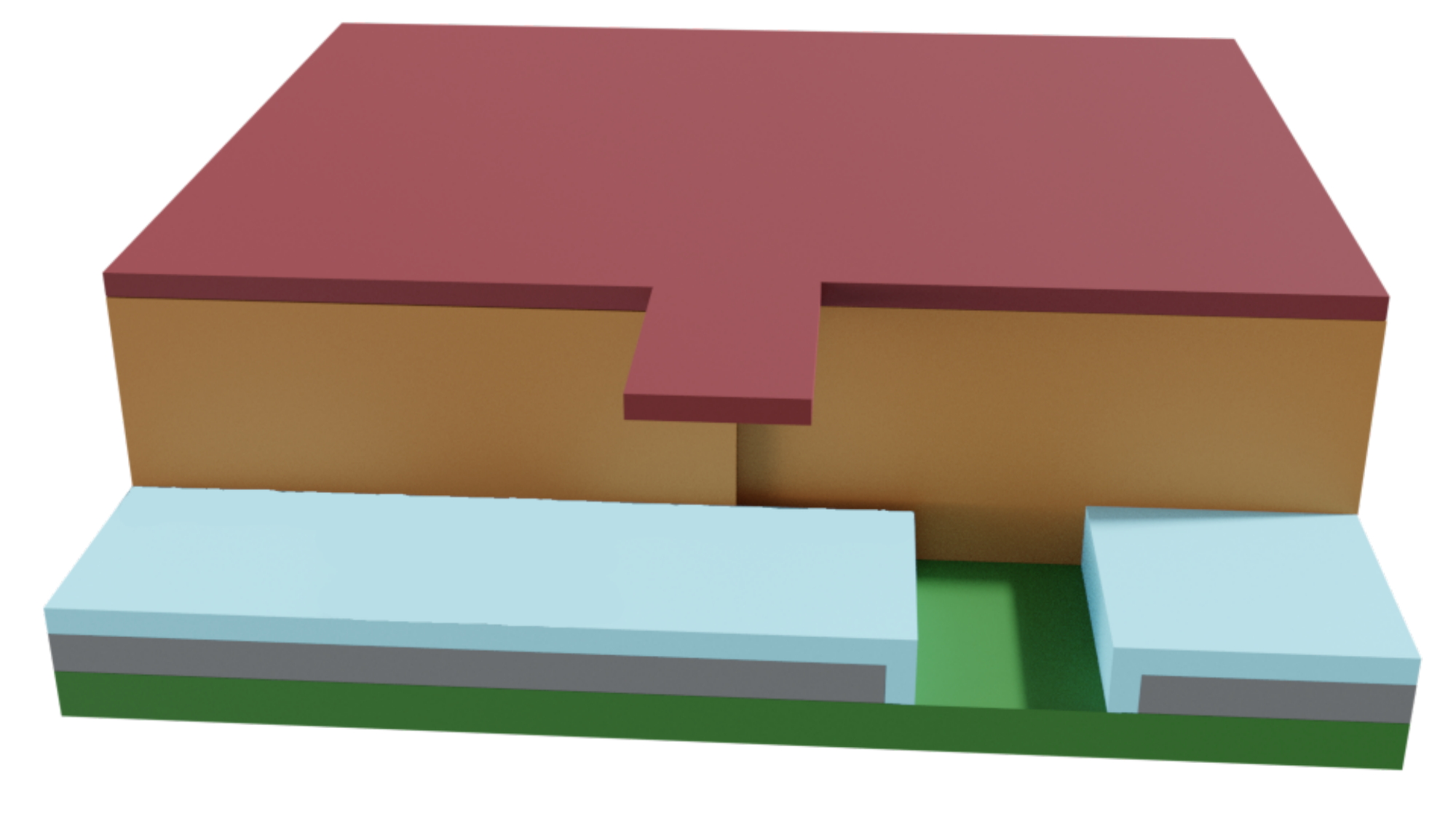}}
\hfill
\subfloat[Deposition of \SI{30}{\nano \metre} aluminum at \SI{1}{\nano \metre \per \second} and an angle of -\ang{19}, forming the top electrode.]{\includegraphics[width = 0.33\textwidth]{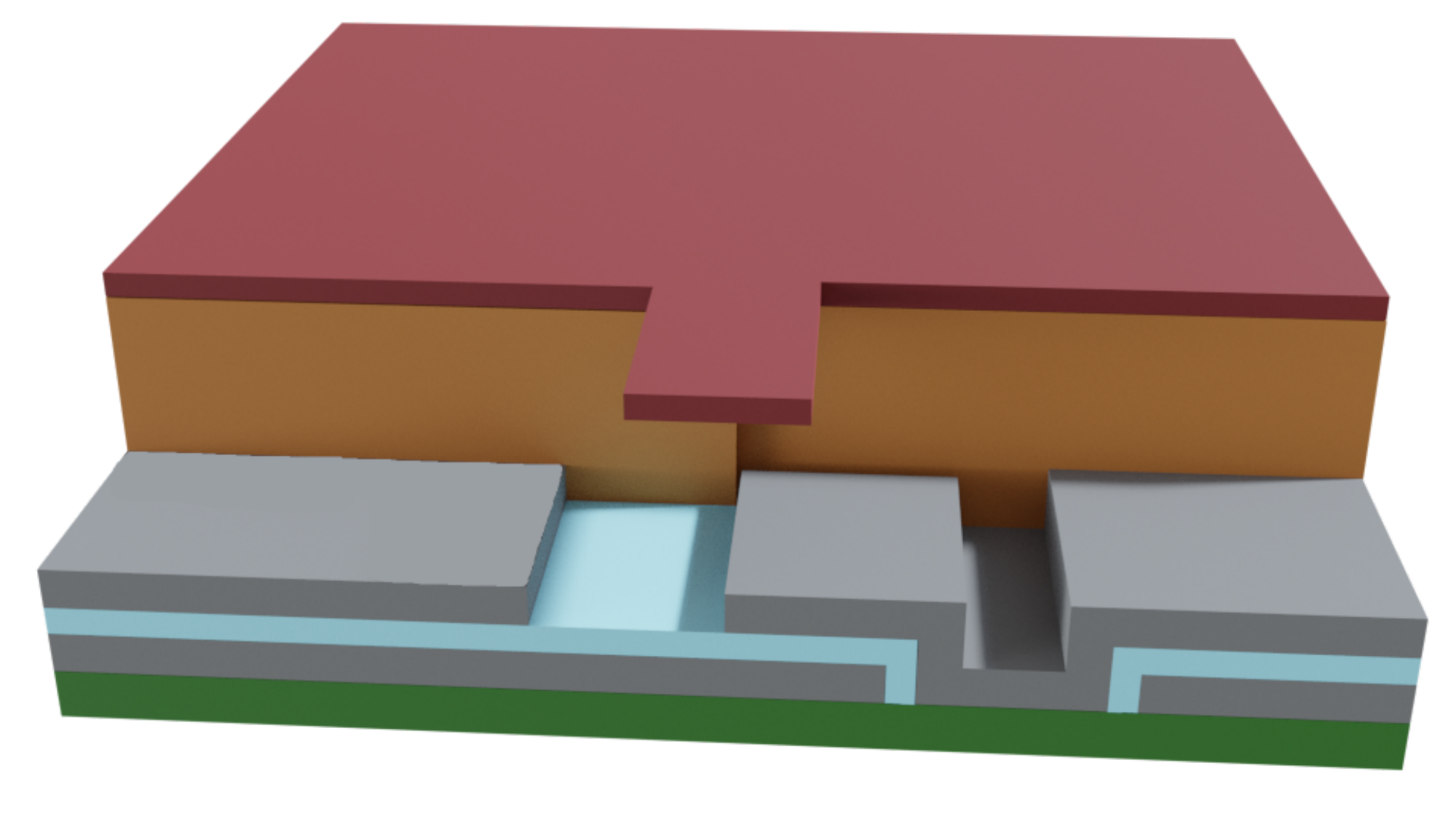}}
\hfill
\subfloat[Argon-milling at \ang{0} to remove part of the oxide layer to make direct contact with the bandage layer.]{\includegraphics[width = 0.33\textwidth]{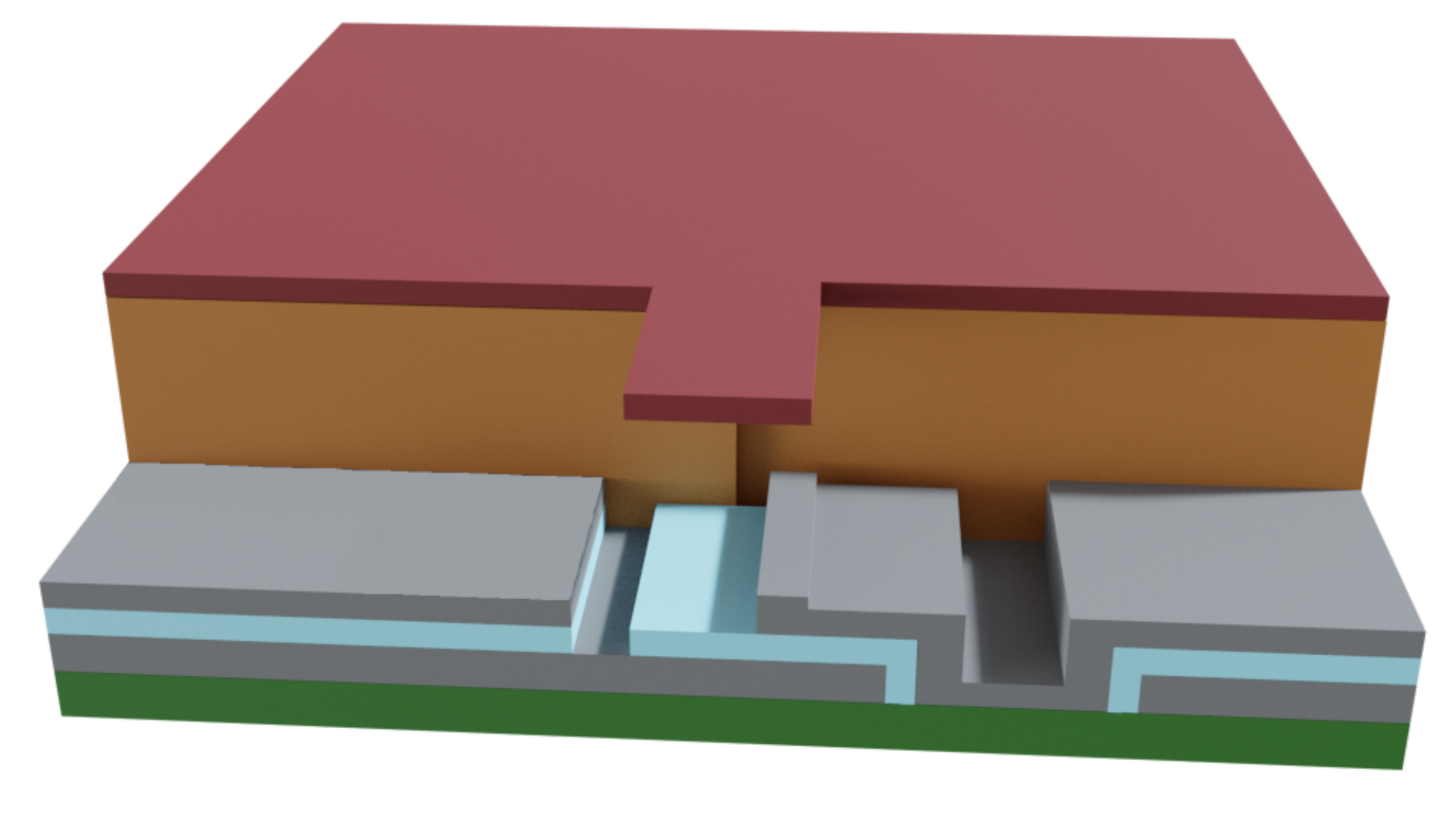}}
\vskip 1em
\subfloat[Deposition of \SI{140}{\nano \metre} aluminum at \SI{1}{\nano \metre \per \second} and an angle of \ang{0}, forming the bandage layer.]{\includegraphics[width = 0.33\textwidth]{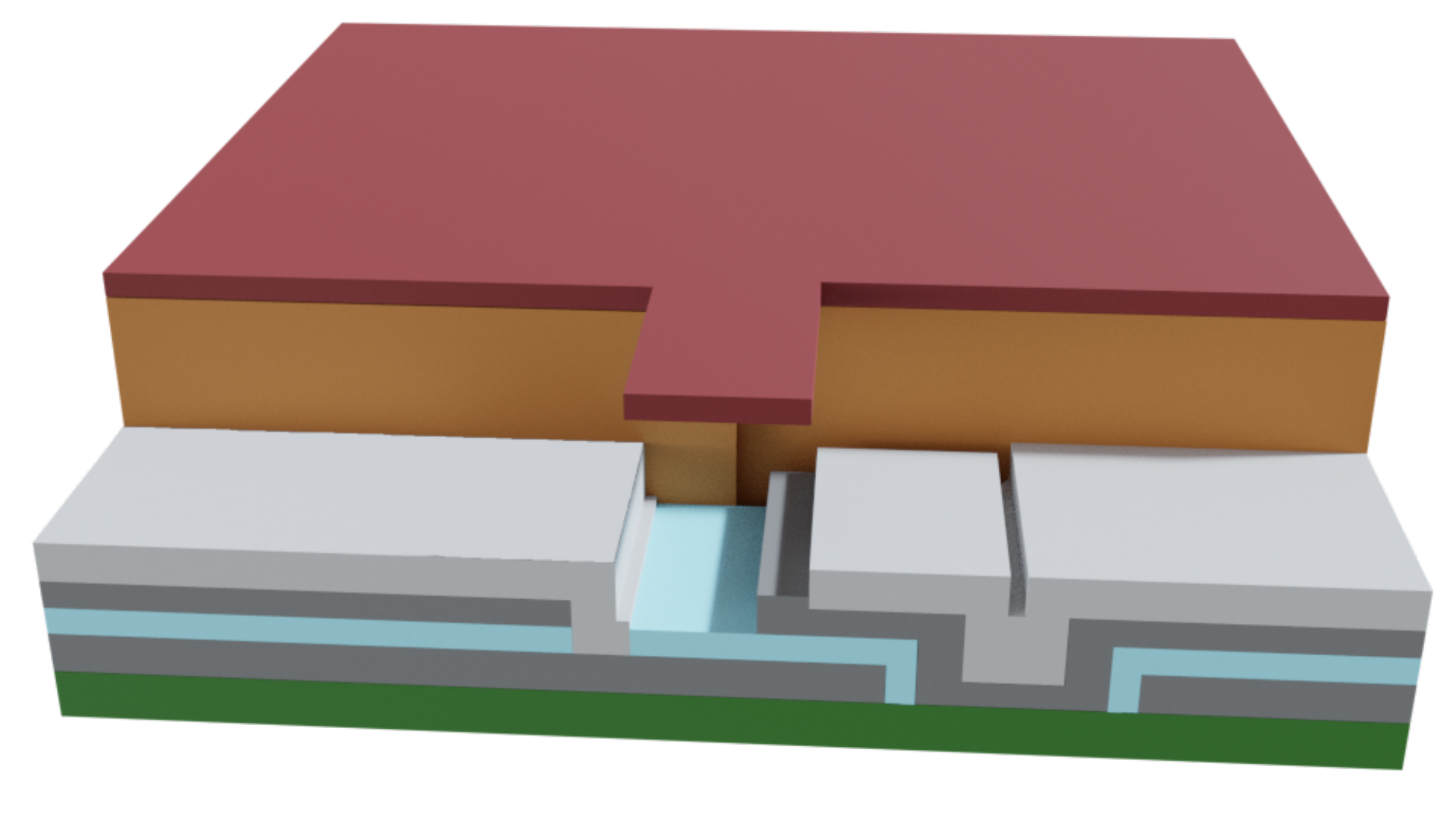}}
\caption{Sketch of the shadow evaporation forming the Josephson junctions. Green: sapphire substrate, light and dark gray: aluminum, blue: aluminum oxide, orange: MMA, red: PMMA. For simplicity, the protective oxidations after top electrode evaporation and bandage layer deposition are omitted.}
\label{fig:shadowevaporation}
\end{figure*}

The starting point of the fabrication is a C-plane oriented, \SI{500}{\micro \metre} thick, 3in. sapphire wafer. In a first step, the wafer is cleaned from organic residuals and other contaminants by means of Piranha solution and an oxygen plasma. It is then installed in a \textit{PLASYS MEB 550 S}, heated to $\SI{200}{\celsius}$ for 2h to remove any excess moisture, and subsequently evaporated with $\SI{100}{\nano \metre}$ aluminum at a rate of $\SI{1}{\nano \metre \per \second}$, a polar angle of $\ang{0}$, and a base pressure of $5 \times 10^{-8}\si{\milli \bar}$ to form the ground plane. Finally, the aluminum layer is passivated by means of a static oxidation at $\SI{30}{\milli \bar}$ for $\SI{10}{\minute}$.

This aluminum ground plane is then structured as discussed in Secs. \ref{sec:des} and \ref{sec:fab} of the main text using optical lithography. For this purpose, \textit{S1805} photoresist is applied and patterned using a mask-aligner and \textit{AZ Developer}. The pattern is subsequently transferred to the aluminum ground plane using an argon-chlorine plasma in an inductively coupled plasma etcher or through wet etching with \textit{TechniEtch Al80} in an ultrasonic bath. Afterward, the resist is removed and the wafer cleaned with DMSO, 2-propanol and an oxygen plasma.

The remaining structures, such as the Josephson junctions, are subsequently deposited using the \textit{in-situ} bandaged Niemeyer-Dolan technique. Therefore, a bilayer of $\SI{250}{\nano \metre}$ \textit{A4} (PMMA) on top of $\SI{900}{\nano \metre}$ \textit{EL-13} (MMA) is applied to the sample and patterned using an electron beam writing system and a mixture of 2-propanol and bidistilled water. Resist residuals in the trenches are removed using an oxygen plasma. Afterward, the Josephson junctions and leads are again deposited in the \textit{PLASYS MEB 550 S} at a base pressure of $5 \times 10^{-8}\si{\milli \bar}$, as illustrated in Fig. \ref{fig:shadowevaporation}. After steps (e) and (g), a static, protective oxidation at $\SI{30}{\milli \bar}$ for $\SI{10}{\minute}$ is performed. Finally, a lift-off is performed using DMSO and the sample is cleaned in 2-propanol, and an ultrasonic bath.

To optimize the fabrication process, a total of nine test structure samples with varying junction areas, barrier thicknesses, and electron beam exposure doses were fabricated and characterized using four-terminal sensing transport measurements at room temperature and scanning electron microscopy. A comprehensive overview of the measurement results can be found in Tab. \ref{tab:transport}. SEM and optical imaging could explain the comparably low yield of samples 541-1, 541-5, 541-2, and the largest junctions of sample 540-1 with a collapse of the Dolan bridge and SQUID loops being disconnected from the probe pads due to wrongly chosen exposure doses. Further losses could be explained by broken junctions and leads due to contamination and wrong handling during fabrication as well as incorrect transport measurements. 

Investigating the mean resistances of the different samples, large interchip fluctuations of $\SI{5.36}{\kilo \ohm}$ between the identically fabricated and designed samples 541-2, 541-4, 541-6, and 541-3 can be seen. The most likely explanation for this behavior would be fluctuations in the junction barrier thickness. At an evaporation rate of $\SI{0.1}{\nano \metre \per \second}$ and a targeted thickness of $\SI{0.1}{\nano \metre}$, the highest precision of the evaporation system could have been exceeded. If this behavior persists for higher targeted thicknesses remains unknown and needs to be investigated. To ensure a good reproducibility and precise control of the resulting qubit frequencies, this issue should be mitigated in future works. A potential solution is to reduce the pressure and time of the first oxidation to enable the deposition of more aluminum in the second evaporation while maintaining the needed barrier thickness. Furthermore, the two oxidations to form the barrier could be replaced by a single strong oxidation, in accordance with Mamin et al. \cite{Mamin_2021}, or a single UV-assisted oxidation as shown by Fritzsch et al. \cite{UV}. Besides these fluctuations in barrier thickness, a further explanation could lie in chip-to-chip fluctuations in the junction areas induced by misaligned e-beam patterns and subsequent misalignment of the shadow evaporation. 

\begin{figure*}[htbp]
\center
\includegraphics[width = \textwidth]{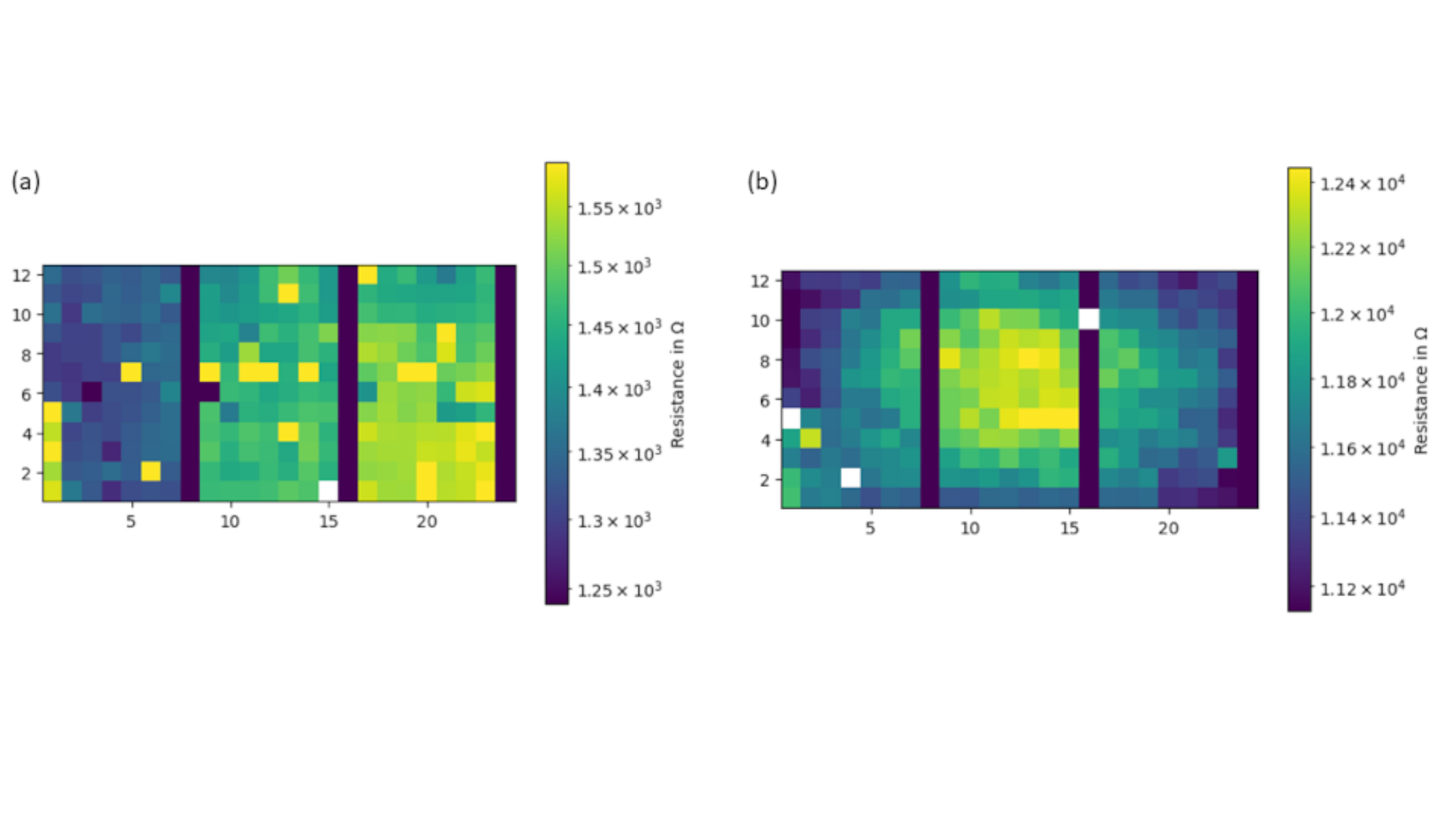}
\caption{Color maps showing the measured spatial distribution of as-fabricated resistances over the respective samples. (a) Sample 540-2 clearly shows a dependence of the SQUID loop resistances on the three different chosen proximity correction exposure doses from left to right. (b) Sample 541-6 shows a radial resistance distribution, indicating a nonuniform e-beam resist application and/or nonuniformly distributed shadow evaporation angles. All mentioned causes result in a variation of the Josephson junction areas.}
\label{fig:resistancevariation}
\end{figure*}

The RSD in resistance of the several samples range from 2.5\% to 10.3\%. As shown in Fig. \ref{fig:resistancevariation}, the large spread can partially be traced back to sweeps in exposure doses, nonuniform resist application and/or nonuniformly distributed shadow evaporation angles resulting in fluctuating junction sizes. A further explanation could be an increased surface roughness due to the double evaporation in combination with a relatively flat evaporation angle of \ang{59} and a low deposition rate \cite{roughness}. Finally, the large, spatially randomly distributed resistance spread of sample 541-3 may be attributed to constrictions in the leads formed by the side trenches.

\begin{figure*}[htbp]
\center
\includegraphics[height = 8cm]{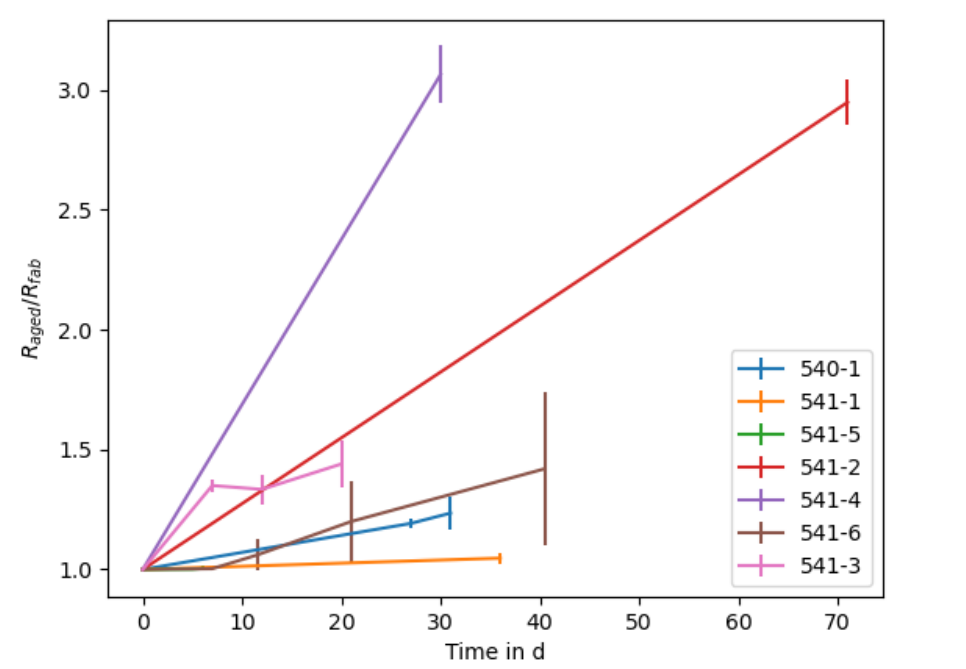}
\caption{Average change in resistance $R_{\mathrm{aged}}/R_{\mathrm{fab}}$ of various samples over time. The effect of junction aging can be seen. Error bars are given by the calculated standard deviation.}
\label{fig:aging}
\end{figure*}

Since resistances can dramatically change within a few days due to junction aging, strongly impacting the resulting critical current density and qubit frequencies, SQUID loop resistances of some samples were monitored over time. As can be seen in Fig. \ref{fig:aging}, junction aging leads to an overall increase in the average sample resistance. This increase greatly varies between different samples. Since all samples were stored in the same way (room temperature, ambient atmospheric conditions, no encapsulation), explanations for these fluctuations can most likely be found in the respective fabrication procedures. More specifically, correlations between the duration of the last cleaning step after lift-off and the degree of junction aging were found.

\begin{table*}[htbp]
\centering
\caption{Results of the statistical analysis of the transport measurements performed directly after fabrication. Here, $R$ and $\sigma_{\mathrm{R}}$ are the mean resistance and relative standard deviation averaged over all successfully measured and fabricated SQUID loops of one sample. $d$ is the thickness of the second deposited aluminum layer during shadow evaporation. $A$ is the total junction area of the SQUID loops. Finally, the number of SQUID loops contributing to the calculated mean resistance and the associated yield are provided as well.}
\label{tab:transport}
    \begin{tabular}{c c c c c c c }
        \toprule
        Sample & $R$ in $\si{\kilo \ohm}$ & $\sigma_{\mathrm{R}}$ in $\%$ & $d$ in $\si{\nano \metre}$ & $A$ in $\si{\micro \metre \squared}$ & Number & Yield in $\%$ \\
        \midrule
         540-2 & 1.433 & 6  & 0  & 2.6  & 238 & 94.4 \\
         540-3 & 17.708 & 4.5 & 1 & 2.6 & 233 & 92.5 \\
         540-1 & 20.892 & 2.9 & 0.8 & 2.6 & 126 & 100 \\
         540-1 & 15.578 & 2.5 & 0.8 & 3.9 & 61 & 96.8 \\
         540-1 & 12.164 & 2.6 & 0.8 & 5.1 & 45 & 71.4 \\
         541-1 & 15.353 & 4 & 0.5 & 2.6 & 136 & 54 \\
         541-5 & 11.445 & 3.6 & 0.3 & 2.6 & 191 & 75.8 \\
         541-2 & 5.413 & 2.6 & 0.1 & 1.97 & 36 & 80 \\
         541-4 & 10.773 & 3.7 & 0.1 & 1.97 & 44 & 97.8\\
         541-6 & 11.766 & 2.8 & 0.1 & 0.99 & 250 & 99.2 \\
         541-3 & 7.227 & 10.3 & 0.1 & 1.97 & 43 & 95.6 \\
        \bottomrule
    \end{tabular}
\end{table*}

\section{Cryogenic setup and $T_{1}$ characterization}\label{app:setup}

\begin{figure*}[htbp]
\center
\includegraphics[height = 8cm]{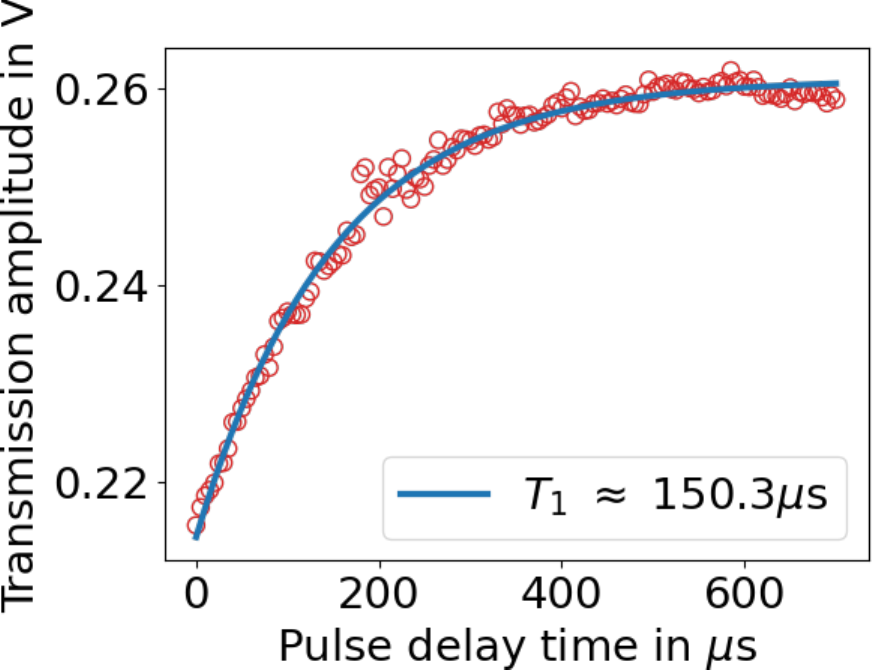}
\caption{Exemplary $T_{1}$ measurement of qubit BGA (red) together with an exponential fit (blue) to extract the respective $T_{1}$ relaxation time.}
\label{fig:single_T1}
\end{figure*}

The cryogenic setup used for qubit characterization is depicted in Fig. \ref{fig:cryogenic}. A single, exemplary $T_{1}$ measurement and respective fit, used to extract the $T_{1}$ times and quality factors given in Sec. \ref{sec:qubit} of the main text, is given in Fig. \ref{fig:single_T1}.

\begin{figure*}[htbp]
\center
\begin{circuitikz}[scale=0.67, transform shape]
\draw (0,0) -- (0,2) to[xgeneric] (2,2) -- (2,0) to[xgeneric, l={$E_{\mathrm{J}}$,$C_{\mathrm{J}}$}] (0,0);
\draw (2,1) to[capacitor, l=$C_{g}$] (4,1);
\draw [<->, thick](2.5,0.5) to[out=300, in=240] (3.5,0.5);
\draw (3,0) node{$g$};
\draw (-0.6,0)node[ground]{} to[inductor] (-0.6,2) -- (-0.6,3) -- (-4.5,3) to[short, -*] (-4.5,4) (-4.5,4.4) to[short, *-] (-4.5,5.5) to[generic] (-4.5,7.5) to[generic] (-4.5,9.5) to[generic] (-4.5,11.5) -- (-4.5,18.3) to[generic] (-4.5,20.3) -- (-4.5,21.1);
\draw [<->, thick](-1.1,1.5) to[out=60, in=120] (0.3,1.5);
\draw (-1.1,2) node{$M$};
\draw (1,1) node{$\Phi$};
\draw (4,-1) to[capacitor, l=$C$] (8,-1) -- (8,3)
to[inductor, l=$L$] (4,3) -- (4,-1);
\draw[rounded corners, dashed, black, thick] (-0.2,-0.8) rectangle (2.2,2.8);
\draw (1,2.5) node{Mergemon};
\draw (6,1) node{CPW resonator};
\draw (-2,25.1) to[short, *-](-2,23.1)to[generic](-2,21.1)-- (-2,21.1) -- (-2,20.3)--(-2,18.3) -- (-2,15.5) to[generic] (-2,13.5) to[generic] (-2,11.5) to[generic] (-2,9.5) to[generic] (-2,7.5) to[generic] (-2,5.5) to[short, -*] (-2,4.4) (-2,4) to[short, *-] (-2,3.5) -- (9.1,3.5) to[short, -*] (9.1,4) (9.1,4.4) to[short, *-] (9.1,6)  (9.1,7) -- (9.1,7.5) to[generic] (9.1,9.5)to[generic] (9.1,11.5)to[generic] (9.1,13.5)--(9.1,14) (9.1,15)--(9.1,15.5) to[generic] (9.1,17.5) -- (9.1,18.3) -- (9.1,18.55) (9.1,20.05) -- (9.1,20.3) -- (9.1,21.35) (9.1,22.85)--(9.1,23.35) (9.1,24.85) to[short, -*] (9.1,25.1);
\draw [<->, thick](6.5,2.5) to[out=30, in=330] (6.5,4);
\draw (7.25,3.25) node{$M$};
\draw (3.55,3.8) node{Transmission line};
\draw (-2.55,2.7) node{Flux bias line};
\draw[rounded corners, dashed, black, ultra thick] (-5,-1.7) rectangle (9.6,4.2);
\draw (-4.8,-2.1) node{Chip};
\draw [ultra thick](-2,4.4) to[out=0, in=0] (-2,4);
\draw [ultra thick](-4.5,4.4) to[out=0, in=0] (-4.5,4);
\draw [ultra thick](9.1,4.4) to[out=0, in=0] (9.1,4);
\draw[rounded corners, brown, ultra thick] (-5.5,-2.4) rectangle (10.1,4.7);
\draw (-4.4,-2.8) node[brown]{Sample holder};
\draw[rounded corners, olive, ultra thick] (-6,-3.1) rectangle (10.6,5.2);
\draw (-4.9,-3.5) node[olive]{$\mu$ metal shield};
\node[rotate=90] at (-2.01,6.5) {IR};
\node[rotate=90] at (-2.01,8.5) {LP};
\node at (-1,8.5) {$\SI{9.5}{\giga \hertz}$};
\node[rotate=90] at (-2.01,10.5) {BP};
\node at (-0.7,10.5) {3.4-$\SI{9.9}{\giga \hertz}$};
\node[rotate=90] at (-2.01,12.5) {$\SI{10}{\deci \bel}$};
\node[rotate=90] at (-2.01,14.5) {$\SI{20}{\deci \bel}$};
\draw (9.1,6.5) node[circulator](C){};
\node at (10.5,6.5) {6-$\SI{11}{\giga \hertz}$};
\node[rotate=90] at (9.09,8.5) {IR};
\node[rotate=90] at (9.09,10.5) {K+L};
\node[rotate=90] at (9.09,12.5) {BP};
\node at (10.4,12.5) {3.4-$\SI{9.9}{\giga \hertz}$};
\draw (9.1,14.5) node[circulator](C){};
\node at (10.7,14.5) {2x 4-$\SI{8}{\giga \hertz}$};
\node[rotate=90] at (9.09,16.5) {K+L};
\node[rotate=90] at (-4.51,6.5) {LP};
\node[rotate=90] at (-4.51,8.5) {LP};
\node[rotate=90] at (-4.51,10.5) {$\SI{20}{\deci \bel}$};
\node at (-3.5,6.5) {$\SI{9.5}{\giga \hertz}$};
\node at (-3.5,8.5) {$\SI{0.8}{\giga \hertz}$};
\draw[rounded corners, dashed, blue, ultra thick] (-6.5,-3.8) rectangle (12,17.5);
\draw (-4.2,-4.2) node[blue]{Mixing chamber (20-$\SI{30}{\milli \kelvin}$)};
\draw[rounded corners, dashed, teal, ultra thick] (-7,-4.5) rectangle (12.5,18);
\draw (-5.6,-4.9) node[teal]{Still (600-$\SI{800}{\milli \kelvin}$)};
\node at (10.5,19.5) {HEMT};
\node at (10.5,19.1) {4-$\SI{8}{\giga \hertz}$};
\node[rotate=90] at (9.09,19.15) {$\SI{30}{\deci \bel}$};
\draw (9.1,20.05)--(8.35,18.55)--(9.85,18.55)--(9.1,20.05);
\node[rotate=90] at (-4.51,19.3) {$\SI{20}{\deci \bel}$};
\draw[rounded corners, dashed, orange, ultra thick] (-7.5,-5.2) rectangle (13,20.3);
\draw (-7.3,-5.6) node[orange]{$\SI{4}{\kelvin}$};
\draw[rounded corners, black, ultra thick] (-8,-5.9) rectangle (13.5,20.8);
\draw (-6.5,-6.3) node[black]{Room temperature};
\draw (-4.5,21.5) node[genericshape, scale = 2]{};
\node at (-4.5,21.5) {AWG};
\draw (9.1,22.85)--(8.35,21.35)--(9.85,21.35)--(9.1,22.85);
\draw (9.1,24.85)--(8.35,23.35)--(9.85,23.35)--(9.1,24.85);
\node[rotate=90] at (9.09,21.95) {$\SI{33}{\deci \bel}$};
\node[rotate=90] at (9.09,23.95) {$\SI{33}{\deci \bel}$};
\draw (3.55,21.5) node[genericshape, scale = 2]{};
\node at (3.55,21.5) {VNA};
\draw (1.95,21.5) to[short, *-] (2.45,21.5) (4.65,21.5) to[short, -*] (5.15,21.5);
\draw (1.95,22.5) to[short, *-] (1.95,23);
\draw (5.15,22.5) to[short, *-] (5.15,24.5);
\draw (1.95,23.5) node[genericsplitter, rotate = 90, scale = 0.5]{};
\draw (1.72,23.9) -- (1.72,24.2) -- (0.18,24.2) -- (0.18,24.5);
\draw (2.18,23.9) -- (2.18,24.5);
\draw (2.18,24.75) node[mixer, scale = 0.5]{};
\draw (2.28,25) -- (2.48,25.5)--(2.48,27);
\draw (2.08,25) -- (1.88,25.5)--(1.88,27);
\draw (2.48,27.55) node[genericshape, rotate = 90]{};
\draw (1.88,27.55) node[genericshape, rotate = 90]{};
\node[rotate=90] at (2.47,27.55) {DAC};
\node[rotate=90] at (1.87,27.55) {DAC};
\draw (5.15,24.75) node[mixer, scale = 0.5]{};
\draw (3.55,25.3) node[genericsplitter, rotate = -90, scale = 0.5]{};
\draw (4.9,24.75) -- (3.78,24.75) -- (3.78,24.9);
\draw (2.43,24.75) -- (3.32,24.75) -- (3.32,24.9);
\draw (5.25,25) -- (5.45,25.5)to[generic](5.45,27);
\draw (5.05,25) -- (4.85,25.5)to[generic](4.85,27);
\node[rotate=90] at (5.44,26.25) {LP};
\node[rotate=90] at (4.84,26.25) {LP};
\draw (5.45,27.55) node[genericshape, rotate = 90]{};
\draw (4.85,27.55) node[genericshape, rotate = 90]{};
\node[rotate=90] at (5.44,27.55) {ADC};
\node[rotate=90] at (4.84,27.55) {ADC};
\draw (4.05,27.55) node[oscillator]{};
\node at (3.55,28.25) {MW2};
\draw (3.55,25.75) to[generic] (3.55,27.05);
\node[rotate=90] at (3.54,26.4) {$\SI{20}{\deci \bel}$};
\node at (5.65,25.2) {I};
\node at (2.68,25.2) {I};
\node at (4.57,25.2) {Q};
\node at (1.6,25.2) {Q};
\draw (1.18,27.55) node[oscillator]{};
\draw (-0.32,27.55) node[genericshape, rotate = 90]{};
\draw (0.18,24.75) node[mixer, scale = 0.5]{};
\draw (0.28,25) -- (0.68,25.5)--(0.68,27.1);
\draw (0.08,25) -- (-0.32,25.5)--(-0.32,27);
\node at (0.68,28.25) {MW1};
\node[rotate=90] at (-0.31,27.55) {DAC};
\draw[rounded corners,dashed, gray, ultra thick] (-1,22.25) rectangle (7,28.5);
\draw (6.5,25.375) node[gray, rotate = 90]{Time domain setup};
\node at (9.1,25.5) {OUT};
\node at (-2,25.5) {IN};
\node at (1.8,21.9) {OUT};
\node at (5.1,21.9) {IN};
\node[rotate=90] at (-2.01,22.1) {$\SI{40}{\deci \bel}$};
\draw[black, ultra thick] (-8,23.75) rectangle (-3,28.5);
\draw (-7.5,28) node[mixer, scale = 0.5]{};
\node at (-6.5,28) {Mixer};
\draw (-7.25,27.25) node[oscillator, scale = 0.5]{};
\node at (-5.43,27.25) {Microwave source};
\draw (-7.5,26.5) node[genericsplitter,rotate = 180, scale = 0.25]{};
\node at (-5.7,26.5) {Power divider};
\draw (-7.5,25.75) node[genericshape, scale = 0.5]{};
\node at (-5.1,25.75) {Attenuator and filter};
\draw (-7.5,25) node[circulator, scale = 0.5](C){};
\node at (-6,25) {Circulator};
\draw (-7.75,24.5)--(-7.75,24)--(-7.25,24.25)--(-7.75,24.5);
\node at (-6.1,24.25) {Amplifier};
\end{circuitikz}
\caption{Schematic of the microwave architecture connecting one cold mergemon qubit to the external readout and control devices at room temperature.}
\label{fig:cryogenic}
\end{figure*}

\section{Purcell limitation}\label{app:purcell}

The $T_{1}$ limitation of a mergemon qubit due to the relaxation into the readout resonator and subsequent dissipation into the transmission line, the so-called Purcell loss, is given by

\begin{equation}
    T_{1}^{\mathrm{Purcell}} = \frac{\Delta^{2}}{g^{2}\kappa}.
\end{equation}

\begin{figure*}[tbh!]
\centering
\subfloat{\includegraphics[width=0.253\textwidth]{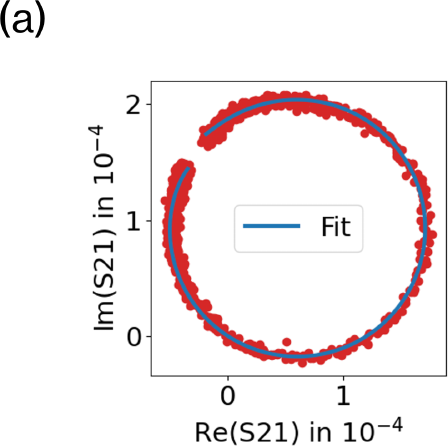}\label{fig:purcellfita}}
\hspace{1cm}
\subfloat{\includegraphics[width=0.325\textwidth]{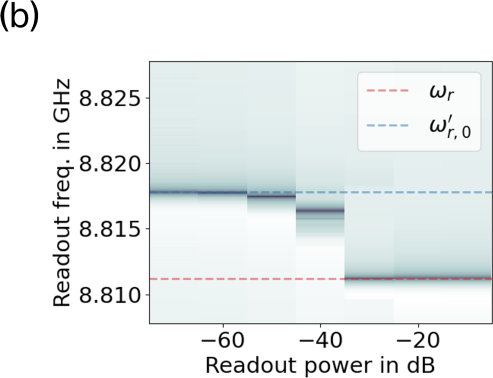}\label{fig:purcellfitb}}
\caption{(a) Complex transmission data (red) of the readout resonator of qubit BGA obtained at zero flux and $\SI{-70}{\deci \bel}$ readout power with a VNA. The circle fit provided by the python package \textit{qkit} and used to extract the loaded quality factor and resonance frequency of the readout resonator is indicated in blue. (b) Power scan (in steps of 10dB) of the readout resonator of qubit BGA obtained at zero flux using a VNA. For readout powers far below the critical photon number $n_{\mathrm{crit}} \approx (\Delta/2g)^2$ the resonator frequency is dispersively shifted by the mergemon qubit being in its ground state to $\omega_{\mathrm{r,0}}'$ (blue dashed line). At high powers above $n_{\mathrm{crit}}$, the resonator approximately approaches its bare frequency $\omega_{\mathrm{r}}$ without the qubit being present (red dashed line).}
\label{fig:purcellfit}
\end{figure*}

Here, g is the coupling strength of the qubit to the resonator, $\Delta = \omega_{\mathrm{q}}-\omega_{\mathrm{r}}$ is the qubit-resonator detuning, $\kappa = \omega_{\mathrm{r}}/Q_{\mathrm{r}}$ is the loss rate of the resonator, and $Q_{\mathrm{r}}$ is the loaded quality factor of the resonator. As shown in Fig. \ref{fig:purcellfita}, the resonator's loaded quality factor $Q_{\mathrm{r}}$ and resonance frequency $\omega_{\mathrm{r}}$ are extracted from a circle fit to the respective transmission data using the \textit{Python} package \textit{qkit} \cite{qkit}. Utilizing the second-order, dispersive Hamiltonian truncated to the first two levels of the transmon \cite{RevModPhys.93.025005}

\begin{equation}
    H_{\mathrm{disp}} \approx \hbar \omega_{\mathrm{r}}''\hat{a}^\dagger\hat{a} + \frac{\hbar \omega_{\mathrm{q}}'}{2}\hat{\sigma}_{\mathrm{z}}+\hbar \chi \hat{a}^\dagger\hat{a} \hat{\sigma}_{\mathrm{z}},
\end{equation}

the coupling strength can be expressed by

\begin{equation}
    g = \sqrt{\frac{(\omega_{\mathrm{r}}-\omega_{\mathrm{r}}')(\Delta-E_{\mathrm{C}}/\hbar)}{1+E_{\mathrm{C}}/(\hbar \Delta)}}.
\end{equation}

Here, $\omega_{\mathrm{r}}$ is the bare resonator frequency, $\omega_{\mathrm{r}}'$ is the displaced resonator frequency due to the qubit being in its ground state, and $E_{\mathrm{C}}$ is the transmon's charging energy. Hence, the coupling strength at zero flux $g_{0}$ can be obtained from the known qubit frequency at zero flux $\omega_{\mathrm{q,0}}$ and charging energy (see Tab. \ref{tab:qubitparameters} of the main text) in combination with resonator power scans as shown in Fig. \ref{fig:purcellfitb}. To assess the Purcell limitation not at zero flux, but at the qubit frequency $\omega_{\mathrm{q,T_{1}}}$ at which respective $T_{1}$ measurements were taken, the coupling strength can be readily scaled down by

\begin{equation}
    g_{\mathrm{T_{1}}} = g_{0}\sqrt{\frac{\omega_{\mathrm{q,T_{1}}}}{\omega_{\mathrm{q,0}}}}.
\end{equation}

All obtained results are listed in Tab. \ref{tab:purcell}.

\begin{table*}[htbp]
\centering
\caption{Results obtained on the Purcell limitation analysis for all mergemon qubits characterized in the main text, except for qubit AB2. Here, $f_{\mathrm{r}}$ is the bare readout resonator frequency, $f_{\mathrm{r,0}}'$ is the dispersively shifted resonator frequency for the qubit being in its ground state at zero flux, $g_{0}/2\pi$ is the qubit-resonator coupling strength at zero flux, $f_{\mathrm{q,T_{1}}}$ is the qubit frequency at which the respective $T_{1}$ measurements provided in Fig. \ref{fig:T1b} of the main text were measured, $\kappa/2\pi$ is the loss rate of the resonator, and $T_{1}^{\mathrm{Purcell}}$ is the resulting $T_{1}$ limitation due to Purcell loss to the readout resonator at the respective qubit frequency $f_{\mathrm{q,T_{1}}}$.}
\label{tab:purcell}
    \begin{tabular}{c c c c c c c}
        \toprule
        Sample & $f_{\mathrm{r}}$ in $\si{\giga \hertz}$ & $f_{\mathrm{r}}-f_{\mathrm{r,0}}'$ in $\si{\mega \hertz}$ & $g_{0}/2\pi$ in $\si{\mega \hertz}$ & $f_{\mathrm{q,T_{1}}}$ in $\si{\giga \hertz}$ & $\kappa/2\pi$ in $\si{\mega \hertz}$ & $T_{1}^{\mathrm{Purcell}}$ in $\si{\micro \second}$\\
        \midrule
        AA & 7.397 & -1.185 & 63 & 4.579 & 2.439 & 137 \\
        AB1 & 7.607 & -0.352 & 35 & 4.481 & 2.530 & 575 \\ 
        AC & 7.257 & -0.142 & 21 & 4.658 & 0.689 & 4000 \\ 
        BF1 & 8.868 & -1.975 & 62 & 5.144 & 1.892 & 447 \\ 
        BF2 & 9.462 & -1.601 & 64 & 5.470 & 1.553 & 555 \\ 
        BF3 & 9.671 & -1.681 & 66 & 5.941 & 1.779 & 375 \\ 
        BF4 & 8.837 & -2.067 & 64 & 5.510 & 0.718 & 822 \\ 
        BGA & 8.811 & -6.593 & 121 & 4.009 & 1.779 & 250 \\ 
        BGB & 8.605 & -5.823 & 108 & 3.846 & 1.179 & 479 \\ 
        \bottomrule
    \end{tabular}
\end{table*}

\section{Junction loss}\label{app:junction}

\begin{figure*}[tbh!]
\centering
\subfloat{\includegraphics[width=0.325\textwidth]{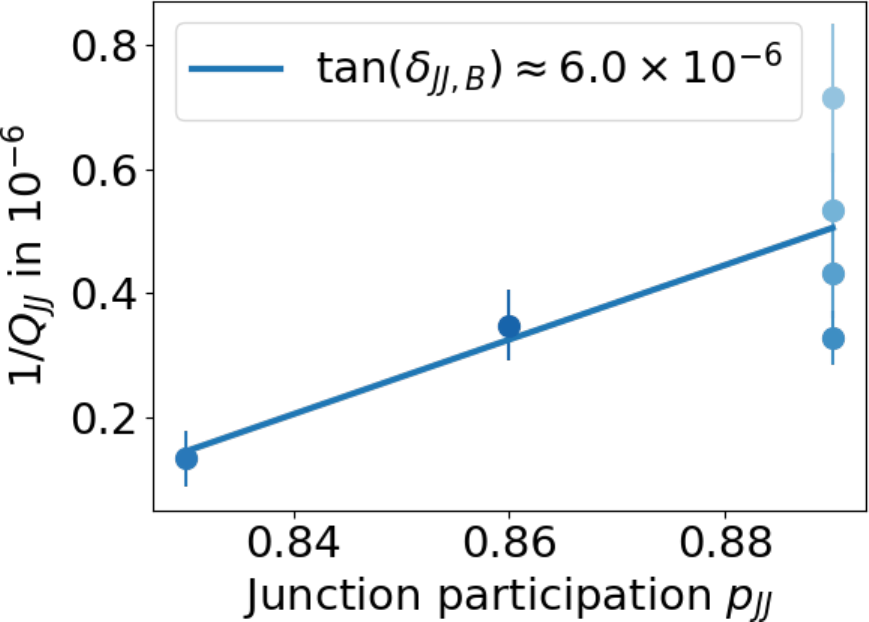}}
\caption{Inverse of the junction loss limited quality factors $Q_{\mathrm{JJ}}$ of qubits following approach B plotted against the simulated junction participation $p_{\mathrm{JJ}}$. A fit to the participation loss model $Q_{\mathrm{JJ}}^{-1} = p_{\mathrm{JJ}} \tan(\delta_{\mathrm{JJ, B}})+C$ is indicated as a blue line.}
\label{fig:junction}
\end{figure*}

Assuming that the $T_{1}$ relaxation times obtained for qubits following design approach B (see Tab. \ref{tab:qubitparameters} of the main text) were dominated by dielectric loss to the Josephson junction (see Sec. \ref{sec:qubit} of the main text for justification), we can neglect surface loss and define the junction loss limited quality factor

\begin{equation}
    Q_{\mathrm{JJ}} \approx \left(Q_{\mathrm{q}}^{-1} - p_{\mathrm{S}} \tan(\delta_{\mathrm{S}}) - Q_{\mathrm{Purcell}}^{-1}\right)^{-1},
\end{equation}

such that $Q_{\mathrm{JJ}}^{-1} = p_{\mathrm{JJ}} \tan(\delta_{\mathrm{JJ, B}})+C$. The resulting values for all mergemon qubits following design approach B are plotted in Fig. \ref{fig:junction} and observed to overall increase with decreasing $p_{\mathrm{JJ}}$, further strengthening the assumption that junction loss is the dominant source of energy relaxation. A fit to the participation loss model (see blue line in Fig. \ref{fig:junction}) gives a relatively low junction loss tangent of $\tan(\delta_{\mathrm{JJ,B}}) = 6 \times 10^{-6}$. We note that, in contrast to the surface loss treatment in Sec. \ref{sec:qubit} of the main text, we needed to include an unphysical, negative offset $C \approx -4.8 \times 10^{-6}$ to correctly fit the data. 

\section{TLS analysis}\label{app:TLS}

\begin{figure*}
\centering
\includegraphics[width = \textwidth]{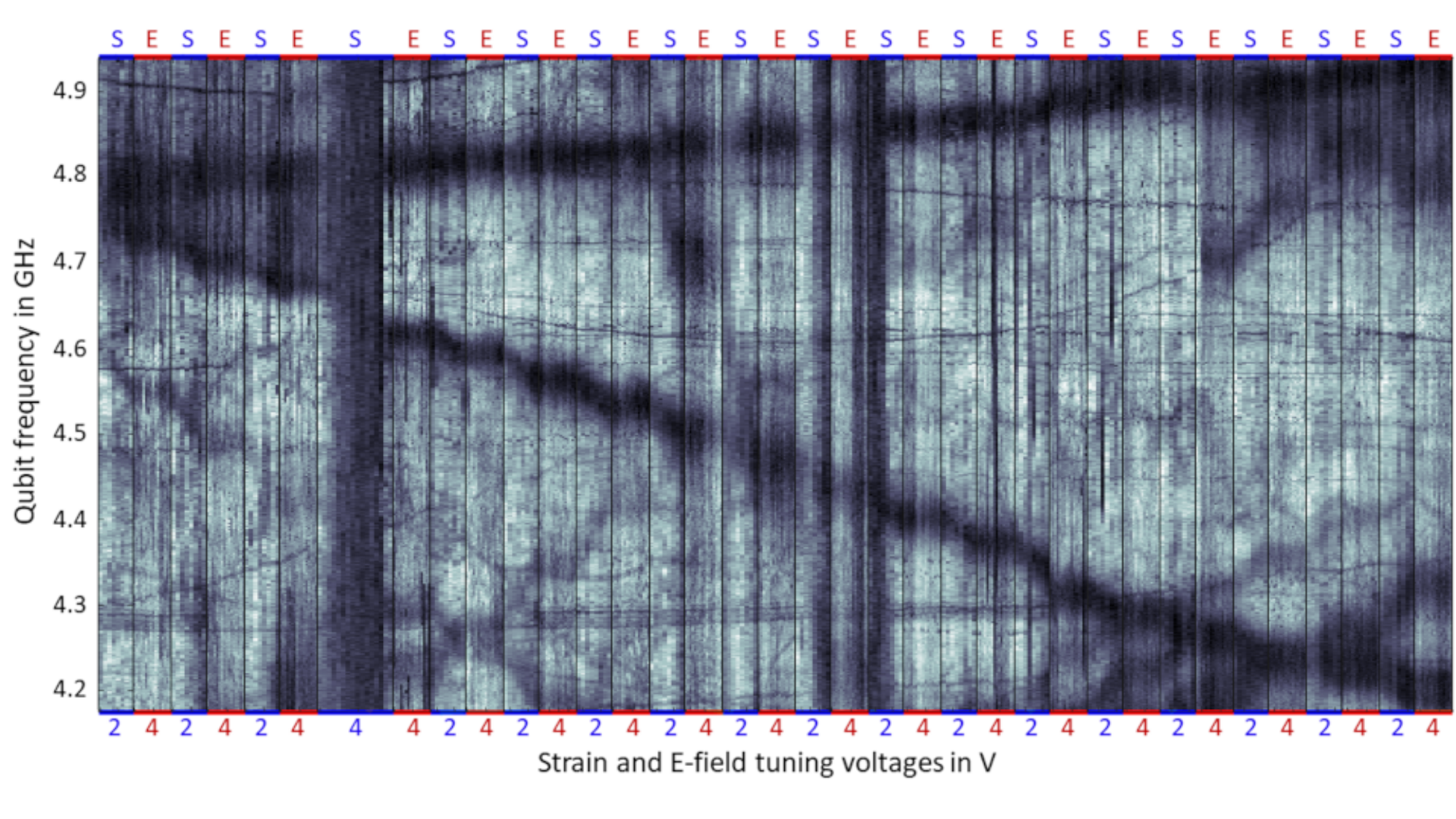}
\caption{TLS spectroscopy of qubit AC. Electric and strain fields were scanned in an alternating fashion of 4 and 2V intervals, respectively. The x-axes are not given in absolute values and are to be taken as additive intervals. The strain and E-field tuning voltage resolutions are 0.25V, while the qubit frequency resolution is $\SI{1}{\mega \hertz}$. The color map provides a measure of the qubit’s $T_{1}$ time with darker regions representing low and lighter regions high values.}
\label{fig:tlsspec}
\end{figure*}

An exemplary result of a strain- and electric-field-dependent swap spectroscopy can be found in Fig. \ref{fig:tlsspec}. For strongly coupled TLS close to their symmetry point $\Delta/\hbar\omega_{\mathrm{TLS}} \approx 1$, the coupling strength between TLS and mergemon qubit is given by

\begin{equation}
    \frac{g}{2 \pi} = \frac{1}{h}\vec{p} \cdot \vec{E},
\end{equation}

with the electric dipole moment of the TLS $\vec{p}$, the electric field $\vec{E}$ induced by the qubit's plasma oscillation at the TLS position, the resonance frequency of the TLS $\omega_{\mathrm{TLS}}$, and the tunneling rate $\Delta$. With the TLS residing inside the Josephson junction and assuming the junction barrier thickness $d_{\mathrm{JJ}}$ to be uniform, the electric field strength at the TLS position can be estimated by

\begin{equation}
    E_{\mathrm{JJ}} = \frac{V_{\mathrm{rms}}}{d_{\mathrm{JJ}}}.
\end{equation}

Here, 

\begin{equation}
    V_{\mathrm{rms}} = \sqrt{\frac{h f_{\mathrm{q}}}{2C_{\Sigma}}}
\end{equation}

is the vacuum voltage fluctuation on the qubit islands at the TLS transition frequency $f_{\mathrm{q}} = f_{\mathrm{TLS}}$ with a total qubit capacitance $C_{\Sigma}$. Furthermore, the qubit-TLS coupling strength can be extracted from coherent swap oscillations by performing a fast Fourier transform (FFT) at the TLS transition frequency

\begin{equation}
    \frac{g}{2\pi} = \frac{f_{\mathrm{FFT}}}{2}.
\end{equation}

The TLS's electric dipole moment parallel to the junction field is hence obtained from

\begin{equation}
    p_{||} = \frac{h}{2} f_{\mathrm{FFT}} d_{\mathrm{JJ}} \sqrt{\frac{2C_{\Sigma}}{hf_{\mathrm{TLS}}}}.
\end{equation}

An estimate for the loss tangent of the junction barrier material can then be given by \cite{Bilmes_2021}

\begin{equation}
    \tan(\delta) = \frac{\pi P_{0} \overline{p}_{||}^{2}}{3\epsilon_{0}\epsilon_{\mathrm{r}}h},
\end{equation}

where $P_{0}$ is the volume junction TLS density and $\overline{p}_{||}$ is the mean electric dipole moment parallel to the junction field. The TLS parameters, obtained from recorded swap oscillations between qubit BGB and seven TLS, can be found in Tab. \ref{tab:swap}.

\begin{table*}[htbp]
\centering
\caption{TLS parameters obtained from swap oscillations between qubit BGB and seven TLS.}
\label{tab:swap}
    \begin{tabular}{c c c c c}
        \toprule
         TLS & $g/2\pi$ in $\si{\mega \hertz}$ & $p_{||}$ in e$\si{\angstrom}$ & $f_{\mathrm{TLS}}$ in $\si{\giga \hertz}$ & $E_{\mathrm{JJ}}$ in $\si{\kilo \volt \per \metre}$\\
        \midrule
         1 & 24.22 & 0.67 & 5.486 & 1.50 \\
         2 & 17.97 & 0.479 & 5.870 & 1.55\\
         3 & 14.07 & 0.41 & 5.018 & 1.43 \\
         4 & 14.07 & 0.38 & 5.598 & 1.51 \\
         5 & 6.64 & 0.19 & 5.330 & 1.48 \\
         6 & 5.08 & 0.134 & 6.048 & 1.57 \\
         7 & 3.91 & 0.11 & 5.168 & 1.45 \\
        \bottomrule
    \end{tabular}
\end{table*}

\clearpage

\bibliography{apssamp}

@article{Bilmes_2020,
   title={Resolving the positions of defects in superconducting quantum bits},
   volume={10},
   ISSN={2045-2322},
   url={http://dx.doi.org/10.1038/s41598-020-59749-y},
   DOI={10.1038/s41598-020-59749-y},
   number={1},
   journal={Scientific Reports},
   publisher={Springer Science and Business Media LLC},
   author={Bilmes, Alexander and Megrant, Anthony and Klimov, Paul and Weiss, Georg and Martinis, John M. and Ustinov, Alexey V. and Lisenfeld, Jürgen},
   year={2020},
   month={02}
}

@article{Lisenfeld_2019,
   title={Electric field spectroscopy of material defects in transmon qubits},
   volume={5},
   ISSN={2056-6387},
   url={http://dx.doi.org/10.1038/s41534-019-0224-1},
   DOI={10.1038/s41534-019-0224-1},
   number={1},
   journal={npj Quantum Information},
   publisher={Springer Science and Business Media LLC},
   author={Lisenfeld, Jürgen and Bilmes, Alexander and Megrant, Anthony and Barends, Rami and Kelly, Julian and Klimov, Paul and Weiss, Georg and Martinis, John M. and Ustinov, Alexey V.},
   year={2019},
   month={11} 
}

@article{Martinis2,
  title = {Decoherence in Josephson Qubits from Dielectric Loss},
  author = {Martinis, John M. and Cooper, K. B. and McDermott, R. and Steffen, Matthias and Ansmann, Markus and Osborn, K. D. and Cicak, K. and Oh, Seongshik and Pappas, D. P. and Simmonds, R. W. and Yu, Clare C.},
  journal = {Phys. Rev. Lett.},
  volume = {95},
  issue = {21},
  pages = {210503},
  numpages = {4},
  year = {2005},
  month = {Nov},
  publisher = {American Physical Society},
  doi = {10.1103/PhysRevLett.95.210503},
  url = {https://link.aps.org/doi/10.1103/PhysRevLett.95.210503}
}

@article{mergemonnotmamain,
  title = {Merged-Element Transmon},
  author = {Zhao, R. and Park, S. and Zhao, T. and Bal, M. and McRae, C.R.H. and Long, J. and Pappas, D.P.},
  journal = {Phys. Rev. Appl.},
  volume = {14},
  issue = {6},
  pages = {064006},
  numpages = {8},
  year = {2020},
  month = {12},
  publisher = {American Physical Society},
  doi = {10.1103/PhysRevApplied.14.064006},
  url = {https://link.aps.org/doi/10.1103/PhysRevApplied.14.064006}
}

@article{Mamin_2021,
   title={Merged-Element Transmons: Design and Qubit Performance},
   volume={16},
   ISSN={2331-7019},
   url={http://dx.doi.org/10.1103/PhysRevApplied.16.024023},
   DOI={10.1103/physrevapplied.16.024023},
   number={2},
   journal={Physical Review Applied},
   publisher={American Physical Society (APS)},
   author={Mamin, H.J. and Huang, E. and Carnevale, S. and Rettner, C.T. and Arellano, N. and Sherwood, M.H. and Kurter, C. and Trimm, B. and Sandberg, M. and Shelby, R.M. and Mueed, M.A. and Madon, B.A. and Pushp, A. and Steffen, M. and Rugar, D.},
   year={2021},
   month={8} 
}

@article{Google1,
    title={Suppressing quantum errors by scaling a surface code logical qubit}, 
    author={Acharya, Rajeev and Aleiner, Igor and Allen, Richard and Andersen, Trond I. and Ansmann, Markus and Arute, Frank and Arya, Kunal and Asfaw, Abraham and Atalaya, Juan and Babbush, Ryan and others},
    journal = {Nature},
    volume = {614},
    issue = {7949},
    pages = {676-681},
    numpages = {6},
    year={2023},
    month = {2},
    publisher = {Springer Science and Business Media LLC},
    doi = {10.1038/s41586-022-05434-1},
    url = {https://doi.org/10.1038/s41586-022-05434-1}
}

@article{Google2,
    title = {Quantum supremacy using a programmable superconducting processor},
    author = {Arute, Frank and Arya, Kunal and Babbush, Ryan and Bacon, Dave and Bardin, Joseph C. and Barends, Rami and Biswas, Rupak and Boixo, Sergio and Brandao, Fernando G. S. L. and Buell, David A. and others},
    doi = {10.1038/s41586-019-1666-5},
    url = {https://doi.org/10.1038/s41586-019-1666-5},
    journal = {Nature},
    volume = {574},
    issue = {7779},
    pages = {505-510},
    numpages = {6},
    year = {2019},
    month = {10},
    publisher = {Springer Science and Business Media LLC}
}

@article{IBM,
    doi = {10.1088/2058-9565/abe519},
    url = {https://dx.doi.org/10.1088/2058-9565/abe519},
    year = {2021},
    month = {3},
    publisher = {IOP Publishing},
    volume = {6},
    number = {2},
    pages = {025020},
    author = {Jurcevic, Petar and Javadi-Abhari, Ali and Bishop, Lev S and Lauer, Isaac and Bogorin, Daniela F and Brink, Markus and Capelluto, Lauren and Günlük, Oktay and Itoko, Toshinari and Kanazawa, Naoki and others},
    title = {Demonstration of quantum volume 64 on a superconducting quantum computing system},
    journal = {Quantum Science and Technology}
}

@article{Koch,
  title = {Charge-insensitive qubit design derived from the Cooper pair box},
  author = {Koch, Jens and Yu, Terri M. and Gambetta, Jay and Houck, A. A. and Schuster, D. I. and Majer, J. and Blais, Alexandre and Devoret, M. H. and Girvin, S. M. and Schoelkopf, R. J.},
  journal = {Phys. Rev. A},
  volume = {76},
  issue = {4},
  pages = {042319},
  numpages = {19},
  year = {2007},
  month = {10},
  publisher = {American Physical Society},
  doi = {10.1103/PhysRevA.76.042319},
  url = {https://link.aps.org/doi/10.1103/PhysRevA.76.042319}
}

@article{Kjaergaard,
    title={Superconducting Qubits: Current State of Play},
    volume={11},
    url={http://dx.doi.org/10.1146/annurev-conmatphys-031119-050605},
    doi={10.1146/annurev-conmatphys-031119-050605},
    number={1},
    journal={Annual Review of Condensed Matter Physics},
    publisher={Annual Reviews},
    author={Kjaergaard, Morten and Schwartz, Mollie E. and Braumüller, Jochen and Krantz, Philip and Wang, Joel I.-J. and Gustavsson, Simon and Oliver, William D.},
    year={2020},
    month={3}, 
    pages={369–395} 
}

@article{Martinis,
    title={Surface loss calculations and design of a superconducting transmon qubit with tapered wiring},
    volume={8},
    url={https://doi.org/10.1038/s41534-022-00530-6},
    doi={10.1038/s41534-022-00530-6},
    journal={npj Quantum Information},
    publisher={Springer Science and Business Media LLC},
    author={Martinis, John M.},
    year={2022},
    month={3}, 
    issue={1} 
}

@article{Bilmes,
doi = {10.1088/1361-6668/ac2a6d},
url = {https://dx.doi.org/10.1088/1361-6668/ac2a6d},
year = {2021},
month = {10},
publisher = {IOP Publishing},
volume = {34},
number = {12},
pages = {125011},
author = {Alexander Bilmes and Alexander K Händel and Serhii Volosheniuk and Alexey V Ustinov and Jürgen Lisenfeld},
title = {In-situ bandaged Josephson junctions for superconducting quantum processors},
journal = {Superconductor Science and Technology},
}

@article{Tantalum1,
    title = {New material platform for superconducting transmon qubits with coherence times exceeding 0.3 milliseconds},
    author = {Place, Alexander P. M. and Rodgers, Lila V. H. and Mundada, Pranav and Smitham, Basil M. and Fitzpatrick, Mattias and Leng, Zhaoqi and Premkumar, Anjali and Bryon, Jacob and Vrajitoarea, Andrei and Sussman, Sara and others},
    doi = {10.1038/s41467-021-22030-5},
    url = {https://doi.org/10.1038/s41467-021-22030-5},
    journal = {Nature Communications},
    volume = {12},
    issue = {1},
    year = {2021},
    month = {3},
    publisher = {Springer Science and Business Media LLC}
}

@article{Tantalum2,
    title = {Surpassing millisecond coherence in on chip superconducting quantum memories by optimizing materials and circuit design},
    author = {Suhas Ganjam and Yanhao Wang and Yao Lu and Archan Banerjee and Chan U Lei and Lev Krayzman and Kim Kisslinger and Chenyu Zhou and Ruoshui Li and Yichen Jia and Mingzhao Liu and Luigi Frunzio and Robert J. Schoelkopf},
    doi = {10.1038/s41467-024-47857-6},
    url = {https://doi.org/10.1038/s41467-024-47857-6},
    journal = {Nature Communications},
    volume = {15},
    issue = {1},
    year = {2024},
    month = {5},
    publisher = {Springer Science and Business Media LLC}
}

@article{Tantalum3,
    title = {Towards practical quantum computers: transmon qubit with a lifetime approaching 0.5 milliseconds},
    author = {Wang, Chenlu and Li, Xuegang and Xu, Huikai and Li, Zhiyuan and Wang, Junhua and Yang, Zhen and Mi, Zhenyu and Liang, Xuehui and Su, Tang and Yang, Chuhong and others},
    doi = {10.1038/s41534-021-00510-2},
    url = {https://doi.org/10.1038/s41534-021-00510-2},
    journal = {npj Quantum Information},
    volume = {8},
    issue = {1},
    year = {2022},
    month = {1},
    publisher = {Springer Science and Business Media LLC}
}

@misc{antennamode,
    author = {O. Rafferty and S. Patel and C. H. Liu and S. Abdullah and C. D. Wilen and D. C. Harrison and R. McDermott},
    title = {Spurious Antenna Modes of the Transmon Qubit},
    eprint = {2103.06803},
    archivePrefix = {arXiv},
    primaryClass = {quant-ph},
    year = {2021},
    month = {3}
}

@article{ambegaokar,
  title = {Tunneling Between Superconductors},
  author = {Ambegaokar, Vinay and Baratoff, Alexis},
  journal = {Phys. Rev. Lett.},
  volume = {10},
  issue = {11},
  pages = {486--489},
  numpages = {0},
  year = {1963},
  month = {6},
  publisher = {American Physical Society},
  doi = {10.1103/PhysRevLett.10.486},
  url = {https://link.aps.org/doi/10.1103/PhysRevLett.10.486}
}

@article{Mueller_2019,
doi = {10.1088/1361-6633/ab3a7e},
url = {https://dx.doi.org/10.1088/1361-6633/ab3a7e},
year = {2019},
month = {oct},
publisher = {IOP Publishing},
volume = {82},
number = {12},
pages = {124501},
author = {Clemens Müller and Jared H Cole and Jürgen Lisenfeld},
title = {Towards understanding two-level-systems in amorphous solids: insights from quantum circuits},
journal = {Reports on Progress in Physics}
}

@article{Goswami_2022,
   title={Towards merged-element transmons using silicon fins: The FinMET},
   volume={121},
   ISSN={1077-3118},
   url={http://dx.doi.org/10.1063/5.0104950},
   DOI={10.1063/5.0104950},
   number={6},
   journal={Applied Physics Letters},
   publisher={AIP Publishing},
   author={Goswami, A. and McFadden, A. P. and Zhao, T. and Inbar, H. and Dong, J. T. and Zhao, R. and McRae, C. R. H. and Simmonds, R. W. and Palmstrøm, C. J. and Pappas, D. P.},
   year={2022},
   month={aug}
}

@article{Pappas_2024,
   title={Alternating-bias assisted annealing of amorphous oxide tunnel junctions},
   volume={5},
   ISSN={2662-4443},
   url={http://dx.doi.org/10.1038/s43246-024-00596-z},
   DOI={10.1038/s43246-024-00596-z},
   number={1},
   journal={Communications Materials},
   publisher={Springer Science and Business Media LLC},
   author={Pappas, David P. and Field, Mark and Kopas, Cameron J. and Howard, Joel A. and Wang, Xiqiao and Lachman, Ella and Oh, Jinsu and Zhou, Lin and Gold, Alysson and Stiehl, Gregory M. and Yadavalli, Kameshwar and Sete, Eyob A. and Bestwick, Andrew and Kramer, Matthew J. and Mutus, Josh Y.},
   year={2024},
   month= {aug} 
}

@article{weeden2025statisticsstronglycoupleddefects,
  title = {Statistics of strongly coupled defects in superconducting qubits},
  author = {S. Weeden and D. C. Harrison and S. Patel and M. Snyder and E. J. Blackwell and G. Spahn and S. Abdullah and Y. Takeda and B. L. T. Plourde and J. M. Martinis and R. McDermott},
  journal = {Phys. Rev. Appl.},
  volume = {25},
  issue = {4},
  pages = {044050},
  numpages = {13},
  year = {2026},
  month = {Apr},
  publisher = {American Physical Society},
  doi = {10.1103/4ssz-6ctb},
  url = {https://link.aps.org/doi/10.1103/4ssz-6ctb}
}

@inproceedings{wang2024precision,
  title={Precision frequency tuning of tunable transmon qubits using alternating-bias assisted annealing},
  author={Xiqiao Wang and Joel Howard and Eyob A. Sete and Greg Stiehl and Cameron Kopas and Stefano Poletto and Xian Wu and Mark Field and Nicholas Sharac and Christopher Eckberg and others},
  booktitle={2024 IEEE International Conference on Quantum Computing and Engineering (QCE)},
  volume={1},
  pages={1315--1323},
  year={2024},
  organization={IEEE}
}

@article{Kreikebaum_2020,
doi = {10.1088/1361-6668/ab8617},
url = {https://dx.doi.org/10.1088/1361-6668/ab8617},
year = {2020},
month = {apr},
publisher = {IOP Publishing},
volume = {33},
number = {6},
pages = {06LT02},
author = {Kreikebaum, J M and O’Brien, K P and Morvan, A and Siddiqi, I},
title = {Improving wafer-scale Josephson junction resistance variation in superconducting quantum coherent circuits},
journal = {Superconductor Science and Technology}
}

@article{Osman_2021,
   title={Simplified Josephson-junction fabrication process for reproducibly high-performance superconducting qubits},
   volume={118},
   ISSN={1077-3118},
   url={http://dx.doi.org/10.1063/5.0037093},
   DOI={10.1063/5.0037093},
   number={6},
   journal={Applied Physics Letters},
   publisher={AIP Publishing},
   author={Osman, A. and Simon, J. and Bengtsson, A. and Kosen, S. and Krantz, P. and P. Lozano, D. and Scigliuzzo, M. and Delsing, P. and Bylander, Jonas and Fadavi Roudsari, A.},
   year={2021},
   month= {feb}
}

@article{UV,
title = {Preparation of Nb/Al–AlOx/Nb Josephson junctions with critical current densities down to 1 A/cm2},
journal = {Physica C: Superconductivity},
volume = {296},
number = {3},
pages = {319-324},
year = {1998},
issn = {0921-4534},
doi = {https://doi.org/10.1016/S0921-4534(97)01829-7},
url = {https://www.sciencedirect.com/science/article/pii/S0921453497018297},
author = {L Fritzsch and H.-J Köhler and F Thrum and G Wende and H.-G Meyer},
}

@misc{roughness,
      title={Improving Josephson junction reproducibility for superconducting quantum circuits: shadow evaporation and oxidation}, 
      author={D. O. Moskalev and E. V. Zikiy and A. A. Pishchimova and D. A. Ezenkova and N. S. Smirnov and A. I. Ivanov and N. D. Korshakov and I. A. Rodionov},
      year={2022},
      eprint={2212.06692},
      archivePrefix={arXiv},
      primaryClass={quant-ph}
}

@article{Zeng_2015,
doi = {10.1088/0022-3727/48/39/395308},
url = {https://dx.doi.org/10.1088/0022-3727/48/39/395308},
year = {2015},
month = {sep},
publisher = {IOP Publishing},
volume = {48},
number = {39},
pages = {395308},
author = {Zeng, L J and Nik, S and Greibe, T and Krantz, P and Wilson, C M and Delsing, P and Olsson, E},
title = {Direct observation of the thickness distribution of ultra thin AlOx barriers in Al/AlOx/Al Josephson junctions},
journal = {Journal of Physics D: Applied Physics}
}

@article{Wang_2015,
   title={Surface participation and dielectric loss in superconducting qubits},
   volume={107},
   ISSN={1077-3118},
   url={http://dx.doi.org/10.1063/1.4934486},
   DOI={10.1063/1.4934486},
   number={16},
   journal={Applied Physics Letters},
   publisher={AIP Publishing},
   author={Wang, C. and Axline, C. and Gao, Y. Y. and Brecht, T. and Chu, Y. and Frunzio, L. and Devoret, M. H. and Schoelkopf, R. J.},
   year={2015},
   month={oct}
}

@article{PhysRevB.77.180502,
  title = {Suppressing charge noise decoherence in superconducting charge qubits},
  author = {Schreier, J. A. and Houck, A. A. and Koch, Jens and Schuster, D. I. and Johnson, B. R. and Chow, J. M. and Gambetta, J. M. and Majer, J. and Frunzio, L. and Devoret, M. H. and Girvin, S. M. and Schoelkopf, R. J.},
  journal = {Phys. Rev. B},
  volume = {77},
  issue = {18},
  pages = {180502},
  numpages = {4},
  year = {2008},
  month = {May},
  publisher = {American Physical Society},
  doi = {10.1103/PhysRevB.77.180502},
  url = {https://link.aps.org/doi/10.1103/PhysRevB.77.180502}
}

@article{Bilmes_2021,
   title={Quantum sensors for microscopic tunneling systems},
   volume={7},
   ISSN={2056-6387},
   url={http://dx.doi.org/10.1038/s41534-020-00359-x},
   DOI={10.1038/s41534-020-00359-x},
   number={1},
   journal={npj Quantum Information},
   publisher={Springer Science and Business Media LLC},
   author={Bilmes, Alexander and Volosheniuk, Serhii and Brehm, Jan David and Ustinov, Alexey V. and Lisenfeld, Jürgen},
   year={2021},
   month={feb} 
}

@article{Molina_Ruiz_2021,
   title={Origin of mechanical and dielectric losses from two-level systems in amorphous silicon},
   volume={5},
   ISSN={2475-9953},
   url={http://dx.doi.org/10.1103/PhysRevMaterials.5.035601},
   DOI={10.1103/physrevmaterials.5.035601},
   number={3},
   journal={Physical Review Materials},
   publisher={American Physical Society (APS)},
   author={Molina-Ruiz, M. and Rosen, Y. J. and Jacks, H. C. and Abernathy, M. R. and Metcalf, T. H. and Liu, X. and DuBois, J. L. and Hellman, F.},
   year={2021},
   month={mar} 
}

@misc{charpentier2025universalscalingmicrowavedissipation,
      title={Universal scaling of microwave dissipation in superconducting circuits}, 
      author={Thibault Charpentier and Anton Khvalyuk and Lev Ioffe and Mikhail Feigel'man and Nicolas Roch and Benjamin Sacépé},
      year={2025},
      eprint={2507.08953},
      archivePrefix={arXiv},
      primaryClass={cond-mat.mes-hall},
      url={https://arxiv.org/abs/2507.08953}, 
}

@article{Google3,
   title={Quantum error correction below the surface code threshold},
   volume={638},
   ISSN={1476-4687},
   url={http://dx.doi.org/10.1038/s41586-024-08449-y},
   DOI={10.1038/s41586-024-08449-y},
   number={8052},
   journal={Nature},
   publisher={Springer Science and Business Media LLC},
   author={Acharya, Rajeev and Abanin, Dmitry A. and Aghababaie-Beni, Laleh and Aleiner, Igor and Andersen, Trond I. and Ansmann, Markus and Arute, Frank and Arya, Kunal and Asfaw, Abraham and Astrakhantsev, Nikita and others},
   year={2024},
   month={dec},
   pages={920–-926}
}

@article{PhysRevApplied.23.044054,
  title = {Mitigating losses of superconducting qubits strongly coupled to defect modes},
  author = {Colao Zanuz, Dante and Ficheux, Quentin and Michaud, Laurent and Orekhov, Alexei and Hanke, Kilian and Flasby, Alexander and Bahrami Panah, Mohsen and Norris, Graham J. and Kerschbaum, Michael and Remm, Ants and others},
  journal = {Phys. Rev. Appl.},
  volume = {23},
  issue = {4},
  pages = {044054},
  numpages = {13},
  year = {2025},
  month = {Apr},
  publisher = {American Physical Society},
  doi = {10.1103/PhysRevApplied.23.044054},
  url = {https://link.aps.org/doi/10.1103/PhysRevApplied.23.044054}
}

@article{lisenfeld2026mapping,
      title={Mapping the positions of Two-Level-Systems on the surface of a superconducting transmon qubit}, 
      author={Jürgen Lisenfeld and Alexander K. Händel and Etienne Daum and Benedikt Berlitz and Alexander Bilmes and Alexey V. Ustinov},
      journal={npj Quantum Information},
      year={2026},
      volume   = {12},
      number   = {1},
      pages    = {80},
      url={https://doi.org/10.1038/s41534-026-01272-5},
      doi={10.1038/s41534-026-01272-5},
}

@article{RevModPhys.93.025005,
  title = {Circuit quantum electrodynamics},
  author = {Blais, Alexandre and Grimsmo, Arne L. and Girvin, S. M. and Wallraff, Andreas},
  journal = {Rev. Mod. Phys.},
  volume = {93},
  issue = {2},
  pages = {025005},
  numpages = {72},
  year = {2021},
  month = {May},
  publisher = {American Physical Society},
  doi = {10.1103/RevModPhys.93.025005},
  url = {https://link.aps.org/doi/10.1103/RevModPhys.93.025005}
}

@article{Wenner_2011,
   title={Surface loss simulations of superconducting coplanar waveguide resonators},
   volume={99},
   ISSN={1077-3118},
   url={http://dx.doi.org/10.1063/1.3637047},
   DOI={10.1063/1.3637047},
   number={11},
   journal={Applied Physics Letters},
   publisher={AIP Publishing},
   author={Wenner, J. and Barends, R. and Bialczak, R. C. and Chen, Yu and Kelly, J. and Lucero, Erik and Mariantoni, Matteo and Megrant, A. and O’Malley, P. J. J. and Sank, D. and others},
   year={2011},
   month={Sept} }

@article{smirnov2023wiringsurfacelosssuperconducting,
      title={Wiring surface loss of a superconducting transmon qubit}, 
      author={Nikita S. Smirnov and Elizaveta A. Krivko and Anastasiya A. Solovieva and Anton I. Ivanov and Ilya A. Rodionov},
      year={2023},
      volume  = {14},
      number  = {1},
      pages   = {7326},
      journal={Scientific Reports},
      url={https://doi.org/10.1038/s41598-024-57248-y},
      doi={10.1038/s41598-024-57248-y}
}

@misc{hedrick2026quantifyingsurfacelossessuperconducting,
      title={Quantifying surface losses in superconducting aluminum microwave resonators}, 
      author={Elizabeth Hedrick and Faranak Bahrami and Alexander C. Pakpour-Tabrizi and Atharv Joshi and Q. Rumman Rahman and Ambrose Yang and Ray D. Chang and Matthew P. Bland and Apoorv Jindal and Guangming Cheng and Nan Yao and Robert J. Cava and Andrew A. Houck and Nathalie P. de Leon},
      year={2026},
      eprint={2603.13183},
      archivePrefix={arXiv},
      primaryClass={quant-ph},
      url={https://arxiv.org/abs/2603.13183}, 
}

@article{Read_2023,
   title={Precision Measurement of the Microwave Dielectric Loss of Sapphire in the Quantum Regime with Parts-per-Billion Sensitivity},
   volume={19},
   ISSN={2331-7019},
   url={http://dx.doi.org/10.1103/PhysRevApplied.19.034064},
   DOI={10.1103/physrevapplied.19.034064},
   number={3},
   journal={Physical Review Applied},
   publisher={American Physical Society (APS)},
   author={Read, Alexander P. and Chapman, Benjamin J. and Lei, Chan U and Curtis, Jacob C. and Ganjam, Suhas and Krayzman, Lev and Frunzio, Luigi and Schoelkopf, Robert J.},
   year={2023},
   month=Mar }

@article{Cho_2023,
   title={Simulating noise on a quantum processor: interactions between a qubit and resonant two-level system bath},
   volume={8},
   ISSN={2058-9565},
   url={http://dx.doi.org/10.1088/2058-9565/acf685},
   DOI={10.1088/2058-9565/acf685},
   number={4},
   journal={Quantum Science and Technology},
   publisher={IOP Publishing},
   author={Cho, Yujin and Jasrasaria, Dipti and Ray, Keith G and Tennant, Daniel M and Lordi, Vincenzo and L DuBois, Jonathan and Rosen, Yaniv J},
   year={2023},
   month={Sept}, pages={045023} }

@software{qkit,
author = {qkitgroup and {KIT PHI}},
license = {GPU-2.0},
month = dec,
title = {{Qkit - a quantum measurement suite in python}},
url = {https://github.com/qkitgroup/qkit},
version = {0.4.11},
year = {2024}
}

@misc{kim2026tunablemodelesshybridizationfreecrosskerr,
      title={A Tunable, Modeless, and Hybridization-free Cross-Kerr Coupler for Miniaturized Superconducting Qubits}, 
      author={Gihwan Kim and Andreas Butler and Oskar Painter},
      year={2026},
      eprint={2602.03186},
      archivePrefix={arXiv},
      primaryClass={quant-ph},
      url={https://arxiv.org/abs/2602.03186}, 
}

\end{document}